\documentclass[nofootinbib, reprint, preprintnumbers, amsmath, amssymb, aps, prx, superscriptaddress, floatfix]{revtex4-2}
\usepackage[utf8]{inputenc}
\usepackage{mathtools}
\usepackage{graphicx}
\usepackage{dcolumn}
\usepackage{bm}
\usepackage{physics}
\usepackage{bm}
\usepackage{color}
\usepackage{braket}
\usepackage[normalem]{ulem}
\usepackage{enumerate}
\usepackage{enumitem}
\usepackage[colorlinks=true,linkcolor=blue,citecolor=blue,urlcolor =blue]{hyperref}
\usepackage{mathrsfs}
\usepackage{bbm}

\usepackage{comment}
\usepackage{lipsum}

\newcommand{\argmin}{\mathop{\rm arg~min}\limits}
\newcommand{\argmax}{\mathop{\rm arg~max}\limits}

\begin{document}
\title{How to experimentally evaluate the adiabatic condition for quantum annealing}
\author{Yuichiro Mori}
\email{mori-yuichiro.9302@aist.go.jp}
\affiliation{Global Research and Development Center for Business by Quantum-AI Technology (G-QuAT), National Institute of Advanced Industrial Science and Technology (AIST), 1-1-1, Umezono, Tsukuba, Ibaraki 305-8568, Japan}%

\author{Shiro Kawabata}
\email{s-kawabata@aist.go.jp}
\affiliation{Global Research and Development Center for Business by Quantum-AI Technology (G-QuAT), National Institute of Advanced Industrial Science and Technology (AIST), 1-1-1, Umezono, Tsukuba, Ibaraki 305-8568, Japan}%
\affiliation{NEC-AIST Quantum Technology Cooperative Research Laboratory,
National Institute of Advanced Industrial Science and Technology (AIST), Tsukuba, Ibaraki 305-8568, Japan}

\author{Yuichiro Matsuzaki}
\email[Present E-mail address:]{ymatsuzaki872@g.chuo-u.ac.jp}
\affiliation{Global Research and Development Center for Business by Quantum-AI Technology (G-QuAT), National Institute of Advanced Industrial Science and Technology (AIST), 1-1-1, Umezono, Tsukuba, Ibaraki 305-8568, Japan}%
\affiliation{NEC-AIST Quantum Technology Cooperative Research Laboratory,
National Institute of Advanced Industrial Science and Technology (AIST), Tsukuba, Ibaraki 305-8568, Japan}

\date{\today}

\begin{abstract}
We propose an experimental method for evaluating the adiabatic condition during quantum annealing (QA), which will be essential for solving practical problems.
The adiabatic condition consists of the transition matrix element and the energy gap, and our method simultaneously provides information about these components without diagonalizing the Hamiltonian. The key idea is to measure the power spectrum of a time domain signal by adding an oscillating field during QA, and we can estimate the values of the transition matrix element and energy gap from the measurement output.
Our results provides a powerful experimental basis for analyzing the performance of QA.
\end{abstract}
\maketitle

\section{Introduction}
\label{sec:introduction}
The adiabatic theorem is a crucial result in quantum mechanics, first introduced by Ehrenfest in 1916~\cite{Ehrenfest_1916}. Later, Born and Fock proved a more modern version of the theorem in 1928. The theorem states that if an initial state is prepared in the ground state of the Hamiltonian, it will remain in the ground state as long as the change in the Hamiltonian is slow enough. Since Born and Fock's proof in 1928, there have been numerous studies that have improved and expanded the theorem, including more rigorous formulations \cite{KatoT_1950_JPSJ} and extensions to open systems \cite{Sarandy_2005, Venuti_2016PRA, Dodin_Brumer_PRXQ2021}.

An essential application of this theorem is quantum annealing (QA). This was originally proposed by Apolloni \textit{et al}. in 1989~\cite{Apolloni_1989_stp}. The original proposal aimed to improve the simulated annealing utilizing the quantum effects of tunneling. However, an alternative approach was subsequently presented \cite{Kadowaki_1998_pre,farhi2000quantum}, where the Hamiltonian changes over time.
In this approach, a ground state of the transverse-field Hamiltonian is prepared, and the Hamiltonian is gradually changed to the target problem Hamiltonian. The adiabatic theorem guarantees that if the alteration of the Hamiltonian  is gradual enough, the final state will be the ground state of the problem Hamiltonian.

QA has been intensively studied from various viewpoints, including the computational speed \cite{Somma_2012PRL, Muthukrishnan_2016PRX, Hastings_2021QUANTUM}, implementation methods \cite{Imoto_seki_2022, Miyazaki_2022PRA}, and algorithms \cite{Roland_2002PRA, Chang_2022PRXQ,Schiffer_2022_PRXQ}.
The commercial use of QA machines was pioneered by D-Wave Systems Inc. Accordingly, proposals for their use in research and applications in various fields have arisen, including examples in quantum chemistry \cite{Babbush_2014_scirep, Teplukhin_2020SciRep}, machine learning \cite{Benedetti_2017_prx, Prasanna_2021SciRep}, and high-energy physics \cite{Mott_2017_nature}.

One of the problems in QA is that there is no known efficient method for checking whether the adiabaticity is satisfied or not. In principle, if we can diagonalize the Hamiltonian, we can use an approximate version of the adiabatic conditions are given as follows~\cite{Childs_2001PRA, Morita_2008JMP, Albash_2018, Hauke_RPP2020}:
\begin{align}
     \frac{|\braket{m(s)|\dot{\mathcal{H}}(s)|0(s)}|}{|E_{m}(s)-E_{0}(s)|^{2}}\ll T_{\rm ann} \label{eq:adiabatic_criterion}
\end{align}
for all $s$ and $m$, where $T_{\rm ann}$ 
denotes the annealing time, 
$s=t/T_{\rm ann}$ 
denotes the time normalized by $T_{\rm ann}$, 
$t$ denotes the time,
$\ket{m(s)}$ ($\ket{0(s)}$) denotes the $m$-th excited (ground) state,  $\dot{\mathcal{H}}(s)$ denotes the $s$ derivative of the instantaneous Hamiltonian at a time $s$ and $E_m(s)$ ($E_0(s)$) denote the eigenenergies of the $m$-th excited (ground) state (see Appendix~\ref{sec:adiaba-condi}). Throughout this paper, we consider a dimensionless time $s$ normalized by $T_\mathrm{ann}$.
These conditions are obtained by an argument that considers only the first order perturbation expansion and neglects higher order terms~\cite{Jansen_2007JMP},
and so are not mathematically rigorous. In particular,   conditions~\eqref{eq:adiabatic_criterion} are not known to be sufficient for adiabaticity. However, when the interest is in the qualitative properties of the computation time, these conditions are widely used, and so we adopt them as the adiabatic conditions in our paper.

In the case of applying QA to practical problems, it is unworkable to diagonalize the Hamiltonian with using a classical computer. Consequently, we cannot directly apply the adidabatic conditions \eqref{eq:adiabatic_criterion} to check whether the dynamics is adiabatic or not. Experimental methods have been proposed to measure the energy gap \cite{Matsuzaki_2021JJAP, russo2021evaluating, Schiffer_2022_PRXQ}, which corresponds to the denominator in Eq.~\eqref{eq:adiabatic_criterion}. However, to our knowledge, no studies have been conducted to measure the numerator of the adiabatic condition \eqref{eq:adiabatic_criterion}, \textit{i.e.}, the size of the transition matrix element of the time derivative of the Hamiltonian.

In this paper, we propose a method for simultaneously measuring the numerator and denominator of Eq.~\eqref{eq:adiabatic_criterion}. This method involves utilizing an oscillating field during quantum annealing to induce a Rabi oscillation between the ground and excited states. By performing Fourier transformation on a time domain signal, we obtain a power spectrum and extract relevant information from the data. These steps enable us to evaluate the values of the numerator and denominator of the adiabatic condition~\eqref{eq:adiabatic_criterion}.

The remainder of this paper is organized as follows. In Sec.~\ref{sec:quantum_annealing}, we review QA. In Sec.~\ref{sec:scheme}, we introduce our method for simultaneously measuring the values of the transition matrix element and the energy gap, based on an analytical calculation using some approximations. In Sec.~\ref{sec:Numcal}, we describe numerical simulations (with noise) performed to quantify the performance of our method in realistic cases.
Finally, in Sec.~\ref{sec:conclusion_and_discussion}, we summarize our results and discuss possible directions for future work.

\section{Review of QA}
\label{sec:quantum_annealing}
To review the conventional QA, we consider the following Hamiltonian:
\begin{align}
    \mathcal{H}_{\mathrm{conv}}(s) =f(s) \mathcal{H}_{\mathrm{D}} + (1-f(s)) \mathcal{H}_{\mathrm{P}},\label{def:QA_Hamiltonian_conv}
\end{align}
where $\mathcal{H}_{\mathrm{D}}$ is a driver Hamiltonian, $\mathcal{H}_{\mathrm{P}}$ is a problem Hamiltonian, and $f(s)$ is a schedule function satisfying the condition
\begin{align}
    f(0) = 1,\ f(1) = 0. \label{eq:settings_of_schedulefunc}
\end{align}
Here and in the following we make the choice
\begin{align}
    f(s)&=1-s. \label{eq:ann_linsche}
\end{align}
Due to the condition \eqref{eq:settings_of_schedulefunc}, the Hamiltonian at $s=0$ is the driver Hamiltonian $\mathcal{H}_{\mathrm{D}}$ and the Hamiltonian at $s= 1$ is the problem Hamiltonian. After obtaining a ground state of the driver Hamiltonian, we let the state evolve by the annealing Hamiltonian from $s=0$ to $s=1$. According to the adiabatic theorem, if the annealing time $T_\mathrm{ann}$ is sufficiently large, the state after QA becomes a ground state of the problem Hamiltonian.

\section{Our method for evaluating the adiabatic condition}
\label{sec:scheme}

\begin{figure}
    \centering
    \includegraphics[width = 8cm]{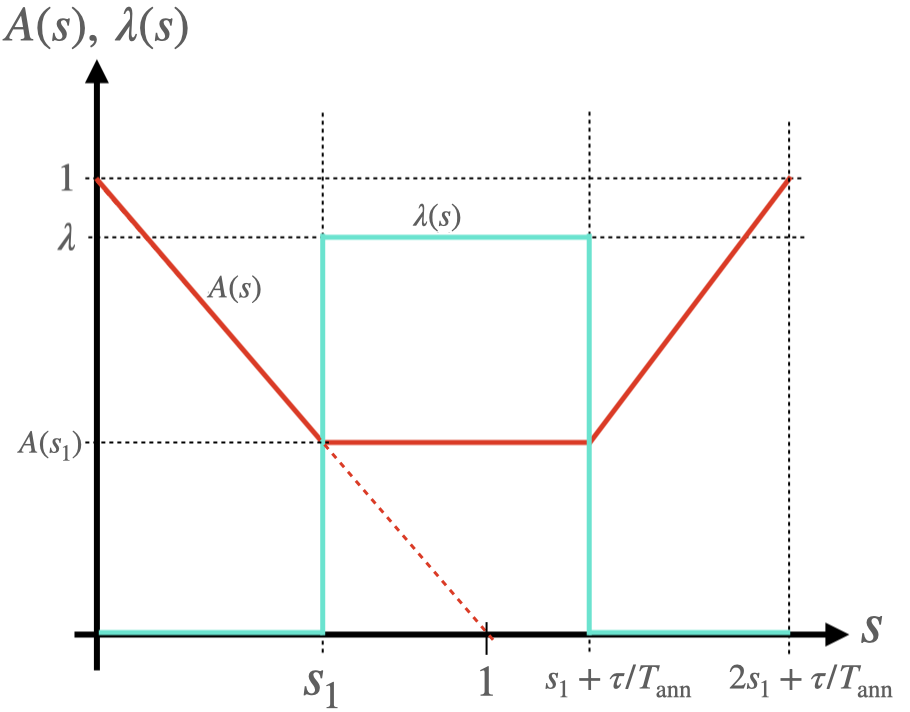}
    \caption{Plot of the scheduling function $A(t)$ and strength of the external driving field $\lambda$ for our protocol. The dotted line shows $f(s)$, which is the scheduling function of the conventional QA.}
    \label{fig:my_label}
\end{figure}

We will now present a technique to experimentally determine the numerator and denominator of the left-hand side of Eq.~\eqref{eq:adiabatic_criterion} for a given time $s_{1}$, using the Hamiltonian defined in Eq.~\eqref{def:QA_Hamiltonian_conv}. In this scenario, the Hamiltonian in Eq.~\eqref{eq:adiabatic_criterion} is the Hamiltonian for quantum annealing $\mathcal{H}_{\mathrm{conv}}$ defined by Eq.~\eqref{def:QA_Hamiltonian_conv}. We introduce the Hamiltonian $\mathcal{H}(s)$, which comprises the driver Hamiltonian $\mathcal{H}_{\rm D}$, the problem Hamiltonian $\mathcal{H}_{\rm P}$, and an external driving Hamiltonian $\mathcal{H}_{\rm ext}(s)$ with strength $\lambda(s)$ and frequency $\omega$ as follows.
\begin{align}
    \mathcal{H}(s) &=\mathcal{H}_{\rm{QA}}(s)
    +\mathcal{H}_{\mathrm{ext}}(s)\label{def:QA_Hamiltonian}
    \\
\mathcal{H}_{\rm{QA}}(s)&= A(s) \mathcal{H}_{\mathrm{D}} + (1-A(s)) \mathcal{H}_{\mathrm{P}}
    \\
    \mathcal{H}_{\mathrm{ext}}(s) &= \lambda (s) \dot{\mathcal{H}}_{\rm{conv}}(s_{1})\cos\left(\omega T_{\rm ann}  (s-s_1)\right)\label{def:external_Hamiltonian}
\end{align}
Here, $A(s)$ is the schedule function that modulates the weight of $\mathcal{H}_{\rm D}$ and $\mathcal{H}_{\rm P}$ in $\mathcal{H}_{\rm QA}(s)$. We plot $A(s)$ and $\lambda(s)$ as functions of time in Fig.~\ref{fig:my_label}, where we note that $A(s)$ satisfies $A(s)=f(s)$ when $0\leq s\leq s_{1}$.

Our experimental protocol proceeds as follows. Firstly, we prepare the ground state of the driver Hamiltonian $\ket{0(s=0)}$. Secondly, we slowly vary the Hamiltonian $H_{\rm QA}(s)$ from $s=0$ to $s=s_{1}$ by setting $\lambda(s) = 0$, allowing the system to evolve under this Hamiltonian adiabatically. Thirdly, at $s=s_{1}$, we introduce a driving term by setting $\lambda(s) = \lambda$ and fixing $A(s)=f(s_{1})$, and we let the system evolve for $s_{1}<s\leq s_{1}+\tau/T_{\rm ann}$. Fourthly, we terminate the driving at $s=s_{1}+\tau/T_{\rm ann}$ by setting $\lambda(s) = 0$, and gradually vary the Hamiltonian from $H_{\rm QA}(s_{1})$ to $H_{\rm D}$ for $s_{1}+\tau/T_{\rm ann} <s\leq 2s_{1}+\tau/T_{\rm ann}$, allowing the system to evolve adiabatically. Finally, we measure the probability of the system occupying the $m$-th excited state $\ket{m(s=0)}$ of the driver Hamiltonian using projective measurements, which we denote as $p_{0,m}(\omega, s_1, \tau)$. We repeat these steps multiple times, varying $\omega, s_{1}$ and $\tau$. We emphasize the importance of adiabaticity during the second and fourth steps, while it is not necessary for the third step.

Let us explain how to realize $\mathcal{H}_{\mathrm{ext}}(s)$ in the third step of the actual experiment. We have
\begin{align}
\mathcal{H}_{\mathrm{ext}}(s)= \lambda
\dot{f}(s_1) \mathcal{H}_{\mathrm{D}}
\cos\left(\omega T_{\rm ann} (s-s_1)\right)\nonumber \\
-\lambda  \dot{f}(s_1) \mathcal{H}_{\mathrm{P}}
\cos\left(\omega T_{\rm ann}(s-s_1)\right).
\end{align}
The driver Hamiltonian and problem Hamiltonian can be decomposed using the Pauli operators as follows:
\begin{align}
    \mathcal{H}_{\mathrm{D}}&=\sum_i h_{i}\mathcal{O}_{i},\\
        \mathcal{H}_{\mathrm{P}}&=\sum_j h'_{j}\mathcal{O}'_{j},
\end{align}
where $\mathcal{O}_{i}$ ($\mathcal{O}'_{j}$) denote the Pauli matrices and $h_{i}$ ($h'_{j}$) denotes a time-independent coefficient.
Hence, we obtain
\begin{align}
\mathcal{H}_{\mathrm{ext}}(s)=\sum _i \lambda
\dot{f}(s_1) h_i \mathcal{O}_{i}
\cos\left(\omega T_{\rm ann} (s-s_1)\right)\nonumber \\
-\sum _{j'}
\lambda
\dot{f}(s_1) h_j' \mathcal{O}_{j'}
\cos\left(\omega T_{\rm ann}(s-s_1)\right).
\end{align}
Thus, if we can temporarily change the coefficient of the Pauli matrices to a cosine function, it is possible to realize the Hamiltonian $\mathcal{H}_{\mathrm{ext}}(s)$. As the problem Hamiltonian usually contains two-body interaction terms, we must change the interaction coupling strength. Such a technique has been developed for superconducting circuits~\cite{bialczak2011fast}.

Here, we describe the dynamics of the system in the third step of our scheme, which is crucial for measuring the adiabatic condition. We begin by describing a simplified scenario in which the dynamics is adiabatic in the second and fourth steps, and we will consider more general cases later. For simplicity, we omit the expression of ``$(s_{1})$'' to mention $\mathcal{H}_{\rm QA}(s_{1})$ or $\dot{\mathcal{H}}_{\rm conv}(s_{1})$ in the remainder of this section. In our proposal, the measurements are performed while sweeping the time period $\tau$; hence, we treat $\tau$ as a variable in the remainder of this section unless mentioned otherwise.

Let us diagonalize $\mathcal{H}_{\mathrm{QA}}$ as follows:
\begin{align}
    \mathcal{H}_{\mathrm{QA}} = \sum_{i}E_{i}\ket{i}\bra{i},\label{eq:diag_qa_hamito}
\end{align}
where $E_i\leq E_j$ is satisfied for $i < j$.
By moving to a rotating frame, we can express the state of the system as follows:
\begin{align}
    \ket{\tilde{\psi}(\tau)}=e^{ir\tau\mathcal{H}_{\mathrm{QA}}}\ket{\psi(\tau)},\label{eq:transformation}
\end{align}
and the Hamiltonian in the rotating frame is expressed as
\begin{align}
    \mathcal{\tilde{H}}(\tau)&= e^{ir\tau\mathcal{H}_{\mathrm{QA}}}\mathcal{H}(\tau)e^{-ir\tau\mathcal{H}_{\mathrm{QA}}}+i\frac{d e^{ir\tau\mathcal{H}_{\mathrm{QA}}}}{d\tau}e^{-ir\tau\mathcal{H}_{\mathrm{QA}}}\nonumber\\
    &=(1-r)\mathcal{H}_{\mathrm{QA}} + e^{ir\tau\mathcal{H}_{\mathrm{QA}}}\mathcal{H}_{\mathrm{ext}}(\tau)e^{-ir\tau\mathcal{H}_{\mathrm{QA}}}.\label{transformed_hamil0}
\end{align}
Note that we set $\hbar = 1$ throughout this paper.
Here, we assume that the transition frequency between the ground state and the $m$-th excited state is close to the frequency of the external driving field. Then, we set
$r$ as the ratio between $|E_{m}-E_{0}|$ and $\omega$ as follows:
\begin{align}
    r=\frac{\omega}{|E_{m}-E_{0}|},\label{eq:definition-s}
\end{align}
where $E_{0}$ denotes the energy of the ground state. The second term in Eq.~\eqref{transformed_hamil0} becomes
\begin{align}
   &e^{ir\tau\mathcal{H}_{\mathrm{QA}}}\mathcal{H}_{\mathrm{ext}}(\tau)e^{-ir\tau\mathcal{H}_{\mathrm{QA}}}\nonumber\\
   &=\lambda \sum_{i, j}\braket{i|\dot{\mathcal{H}}_{\rm conv}|j}e^{ir(E_{i}-E_{j})\tau}\cos{\omega\tau} \ket{i}\bra{j}\label{eq:inthamil_coeff}.
\end{align}
Here, we adopt the rotating wave approximation (RWA)~\cite{Zeuch_2020_aop}.  The coefficient $\ket{i}\bra{j}$ in Eq.~\eqref{eq:inthamil_coeff} includes an oscillatory component:
\begin{align}
    &\quad e^{ir(E_{i}-E_{j})\tau}\cos{\omega\tau} \nonumber\\ &=\frac{1}{2}e^{ir(E_{i}-E_{j})\tau}(e^{i\omega\tau}+e^{-i\omega\tau}) \nonumber\\
    &=\frac{1}{2}e^{i(r(E_{i}-E_{j})-\omega)\tau}+\frac{1}{2}e^{i(r(E_{i}-E_{j})+\omega)\tau}.\label{eq:coeff}
\end{align} 
If $r|E_{i}-E_{j}|= \omega$ is satisfied, one of the terms in Eq.~\eqref{eq:coeff} becomes time-independent while the other term has a high-frequency oscillation. 
Owing to the condition of Eq.~\eqref{eq:definition-s}, we have at least two time-independent terms, $(i,j) = (m,0)$ and $(0,m)$, which remain after 
RWA.
We assume a condition $||E_{m}-E_{0}|-\omega|\ll ||E_{i}-E_{j}|-\omega|$ in neither $(i,j)= (m,0)$ nor  $(i,j) = (0, m)$. Then, all terms except $(i,j) = (0,m)$ and $(i,j)=(m,0)$ are dropped, and the Hamiltonian~\eqref{eq:inthamil_coeff} can be simplified as $\mathcal{H}_{\mathrm{ext},I}=\frac{\lambda}{2}\braket{m|\dot{\mathcal{H}}_{\rm conv}|0}\ket{m}\bra{0} + h.c.$. Therefore, the effective Hamiltonian Eq.~\eqref{transformed_hamil0} can be expressed as
\begin{align}
    \mathcal{H}_{\mathrm{eff}}= \sum_{i} (1-r)E_{i}\ket{i}\bra{i}+\frac{\lambda}{2}\braket{m|\dot{\mathcal{H}}_{\rm conv}|0}\ket{m}\bra{0} + h.c..\label{eq:external-hamil}
\end{align}

These calculations indicate that if the initial state is prepared in a subspace spanned by the ground state and $m$-th excited state, the system's dynamics will be confined to this subspace. Notably, projecting out the states except $\ket{m(s=s_{1})}$ and $\ket{0(s=s_{1})}$ results in an effective Hamiltonian with the same structure as the single-qubit Hamiltonian that induces Rabi oscillations. A known analytical formula that characterizes tha Rabi oscillation without decoherence involves two parameters: detuning and Rabi frequancy, and details of the behavior of Rabi oscillations in a single-qubit system are presented in Appendix~\ref{sec:Rabi-o}. By using this analytical formula, we can fit the data obtained from our method and acquire information about the transition matrix element $|\braket{m|\dot{\mathcal{H}}|0}|$ and the energy gap $E_{\rm m}-E_{0}$.

To observe the oscillation experimentally, we need to construct a projective measurement of $\ket{m}\bra{m}$ in the rotating frame. In our idea, the fourth and fifth steps enable us to construct a projective measurement $\ket{m}\bra{m}$ in the laboratory frame effectively, provided the dynamics in the fourth step is adiabatic. If the state $\ket{\psi(\tau)}$ is an eigenstate of the Hamiltonian $\mathcal{H}_{\rm QA}$, the change in the frame only results in a global phase. Therefore, as long as the second step and fourth step are adiabatically performed, $p_{0,m}(\omega, s_1, \tau)$ is approximately described as follows:
\begin{align}
p_{0,m}(\omega, s_1, \tau) &\simeq
    |\braket{m|e^{-i\tau\mathcal{H}_{\mathrm{eff}}}|0}|^{2} \nonumber\\
    &= \alpha(\omega) (1-\cos\Omega_{\mathrm{ana}}(\omega)\tau),\label{eq:general-Rabi}
\end{align}
where $\alpha(\omega)$ is a function of $\omega$ given by
\begin{align}
    &\alpha(\omega) = \frac{1}{2}\left(\frac{2|\tilde{\lambda}|\left(\Delta-\omega-\sqrt{(\Delta-\omega)^{2}+|\tilde{\lambda}|^{2}}\right)}{\left(\Delta-\omega-\sqrt{(\Delta-\omega)^{2}+|\tilde{\lambda}|^{2}}\right)^{2} + |\tilde{\lambda}|^{2}}\right)^{2},\\
    &\Delta = E_{m} - E_{0},\quad \tilde{\lambda}=\lambda\braket{m|\dot{\mathcal{H}}_{\rm conv}|0},
\end{align}
and $\Omega_{\mathrm{ana}}(\omega)$ is a hyperbolic curve on the $\omega-\Omega$ plane that is represented by
\begin{align}
    \Omega_{\mathrm{ana}}(\omega) = \sqrt{\left(\lambda|\braket{m|\dot{\mathcal{H}}_{\rm conv}|0}|\right)^{2}+(\omega-\Delta)^{2}}.\label{eq:hyperbolic-curve}
\end{align}
We obtain the right-hand side of Eq.~\eqref{eq:general-Rabi}, which is independent of $s_1$, under an assumption that the adiabatic condition is satisfied at the second and fourth steps. However, if there are non-adiabatic transitions, the probability $p_{0,m}$ has a dependence on $s_1$.

In the aforementioned discussion, a ground state of the driver Hamiltonian is assumed to be prepared in the first step, and we perform a projective measurement into the $m$-th excited state in the fifth step. Meanwhile, if we prepare the $k$-th excited state in the first step and perform a projective measurement into the $l$-th excited state in the fifth step, we can obtain the hyperbolic curve as $\Omega_{\mathrm{ana}}^{(k,l)}(\omega) = \sqrt{\left(\lambda|\braket{l|\dot{\mathcal{H}}_{\rm conv}|k}|\right)^{2}+(\omega-\Delta_{kl})^{2}}$, where $\Delta _{kl}=E_{k} - E_{l}$ through similar calculations. The details of these derivations are presented in Appendix~\ref{ubsec:usual-rabio}.

The adiabatic condition described in Eq.~\eqref{eq:adiabatic_criterion} is valid only when we can consider that the effect of the non-adiabatic transitions is weak.
Therefore, throughout our paper (except in the Appendix), we assume that the effect of non-adiabatic transitions is negligible. We will discuss how the non-adiabatic transitions affect the spectroscopic measurements in our methods later.

Let us explain how to specify the values of $|E_m-E_0|$ and $|\braket{m|\dot{\mathcal{H}}_{\rm conv}|0}|$ by using our method. We repeat these by sweeping $\omega$, and we can find an optimal value of $\omega=|E_m-E_0|$ to minimize the frequency of the Rabi oscillation; this corresponds to the energy gap $\Delta$. Furthermore, the Rabi frequency with the optimal $\Omega$ observed in our method corresponds to the numerator in Eq.~\eqref{eq:adiabatic_criterion}. Thus, our estimated transition matrix element $|\braket{m|\dot{\mathcal{H}}|0}|_{\mathrm{est}}$ and our estimated energy gap $\Delta_{\mathrm{est}}$ are given by
\begin{align}
    \lambda |\braket{m|\dot{\mathcal{H}}_{\rm conv}|0}|_{\mathrm{est}}&=\min_{\omega}[\Omega_{\mathrm{exp}}(\omega)],\label{eq:numerator_est}\\
    \Delta_{\mathrm{est}}&=\argmin_{\omega}\left[\Omega_{\mathrm{exp}}(\omega)\right],\label{eq:delta_est}
\end{align}
respectively. Here, $\Omega_{\mathrm{exp}}(\omega)$ is the angular frequency of the Rabi oscillation obtained experimentally, which is analytically considered to be expressed by Eq.~\eqref{eq:hyperbolic-curve}.

In actual experiments, owing to some imperfections, $p_{0,m}(\omega, t_{1}, \tau)$ cannot be fully explained by the analytical formula Eq.~\eqref{eq:general-Rabi}, which was derived under ideal conditions (see Fig.~\ref{fig:Rabi-oscillation-onequbit} in the Appendix). To find the relevant frequency
of $\Omega_{\mathrm{exp}}(\omega)$
in the dynamics, we perform a Fourier transformation and obtain a power spectrum that is defined by
\begin{align}
    P(\omega,s_{1}, \Omega) &= \mathrm{abs}\left[\mathrm{FT}[p_{0,m}(\omega, s_{1}, \tau)]\right],\nonumber\\
    &=\mathrm{abs}\left[\int_{-\infty}^{\infty}d\tau\ p_{0,m}(\omega,s_{1},\tau)\frac{e^{-i\Omega\tau}}{\sqrt{2\pi}}\right].\label{def:power-spectrum}
\end{align}
If $p_{0,m}(\omega,s_{1},\tau)$ is expressed as Eq.~\eqref{eq:general-Rabi}, the power spectrum is given by
\begin{align}
    P(\omega,s_{1},\Omega) = &\alpha(\omega)\delta(\Omega) + \frac{\alpha(\omega)}{2}\delta(\Omega-\Omega_{\mathrm{ana}}(\omega))\nonumber\\ &+\frac{\alpha(\omega)}{2}\delta(\Omega+\Omega_{\mathrm{ana}}(\omega)).\label{eq:ft-ideal-output}
\end{align}
Therefore, in the actual experiment, we define the peak with a positive frequency
in the spectrum as $\Omega_{\mathrm{exp}}(\omega)$, and
we expect to satisfy $\Omega_{\mathrm{exp}}(\omega) \simeq \Omega_{\mathrm{ana}}(\omega)$ in the power spectrum; this allows us to use the formulas of Eqs.~\eqref{eq:numerator_est} and \eqref{eq:delta_est}. Thus, we can estimate the values of the transition matrix element $|\braket{m|\dot{\mathcal{H}}_{\rm conv}|0}|$ and the energy gap $\Delta$ using our method.

\section{Numerical analysis}
\label{sec:Numcal}

We perform numerical simulations to evaluate the effectiveness of our method. In the previous section, we derived an analytical formula under the following assumptions:
\begin{enumerate}[label = \Roman*.]
\item The time evolution is adiabatic in both step 2 and step 4.
\item The rotating wave approximation holds.
\item The time evolution in step 3 only involves the ground state and the $m$-th excited state.
\item There is no decoherence.
\end{enumerate}
However, these assumptions may not be met in actual experiments and we perform numerical simulations to examine the validity of our method under different conditions, as summarized in Table~\ref{tab:simulation-conditions}. 
\begin{table}[hbtp]
    \centering
    
    \begin{tabular}{|c|c|c|c|c|}
    \hline
        Case & Qubit & Adiabaticity of
        & Decoherence & Violated \\
        & Number &Step 2 and 4 & & conditions\\
        \hline
        A & 1 & Complete & None & II \\
        B & 1 & Incomplete & None & I, II\\
        C & 1 & Incomplete & $\checkmark$ & I, II, IV\\
        D & 2 & Complete & None & II, III \\
        E & 2 & Incomplete & None & I, II, III\\
        F & 2 & Incomplete & $\checkmark$ & I, II, III, IV\\
        \hline
    \end{tabular}
    \caption{Cases investigated in this study.
    For cases B, C, E, and F, we consider the effect of non-adiabatic transitions in steps 2 and 4. Meanwhile, for cases C and F, we consider decoherence.}
    \label{tab:simulation-conditions}
\end{table}

Condition I is only satisfied when the process in steps 2 and 4 is completely adiabatic. 
In cases A and D from Table~\ref{tab:simulation-conditions}, we use diagonalization to prepare the ground state of $\mathcal{H}_{\mathrm{QA}}(s_{1})$. In the remaining cases, we solve the time-dependent Schroedinger equation with specific annealing times to prepare the ground state of $\mathcal{H}_{\mathrm{QA}}(s_{1})$.

Condition III is naturally satisfied for a single-qubit system, while it is violated for a system with two or more qubits. Thus, in cases A, B, and C, condition III is satisfied, whereas in cases D, E, and F, condition III is violated.

Condition IV is satisfied if we solve a time-dependent Schroedinger equation of the system, as with cases A, B, D, and E. Meanwhile, we consider the effect of decoherence by solving the master equation in cases C and F.

\subsection{Settings and methods for all cases}
\label{subsec:setandme}
Here, we introduce some conditions that are common throughout our numerical analysis.

\subsubsection{Schedule function}
\label{subsubsec:scfunc}
For the schedule function $A(s)$ in Eq.~\eqref{def:QA_Hamiltonian}, we use
\begin{align}
    A(s) =
    \begin{cases}
    1-s&\ (0\leq s< s_{1}),\\
    1-s_{1}&\ (s_{1}\leq s < s_{1}+\frac{\tau}{T_{\rm ann}}),\\
    s-2s_{1}-\frac{\tau}{T_{\rm ann}}&\ (s_{1}+\frac{\tau}{T_{\rm ann}}\leq s < 2s_{1}+\frac{\tau}{T_{\rm ann}}),
    \end{cases}\label{eq:schedule_function_for_us}
\end{align}
where $T_{\mathrm{ann}}$ is the annealing time.
In actual experiments, this value is typically around $10$ to $100$ $\mu$s, and the typical energy scale of the Hamiltonian is of the order of GHz~\cite{Dwave_datasheet}.

We take the schedule function~\eqref{eq:schedule_function_for_us} as $A(s) = 1-s$ up to $s_1$, and we evaluate the adiabatic condition at time $s_{1}$ according to our method. Hence, for our simulation,
$\dot{\mathcal{H}}_{\rm{QA}}(s_{1})$
is given by
\begin{align}
    \dot{\mathcal{H}}_{\rm conv}(s_{1}) =- \mathcal{H}_{\mathrm{D}} + \mathcal{H}_{\mathrm{P}},\label{eq:ann_ham_d}
\end{align}
for any $t_{1}$.

\subsubsection{Strength $\lambda$}
\label{subsubsec_strength_lambda}

The Rabi frequency can be controlled by changing the strength $\lambda$. 
If the decoherence is negligible, we set $\lambda$ to be as small as possible, because RWA is valid only when the Rabi frequency is much smaller than the energy gap $\Delta$.
Meanwhile, when there is decoherence, the choice of $\lambda$ is not straightforward. As we decrease $\lambda$, the decoherence becomes more relevant and RWA becomes more valid. Therefore, the following condition should be satisfied:
\begin{align}
    \frac{1}{T_{c}|\braket{m|\dot{\mathcal{H}}_{\rm conv}|0}|}\ll \lambda\ll 1,
\end{align}
where $T_{c}$ is the coherence time.
In our simulation, we set $\lambda = 0.05$.

\subsubsection{Time evolution and measurement process}
\label{subsubsec:time_evolution}
In real experiments, we need to consider decoherence. To address this, we use the Gorini--Kossakowski--Sudarshan--Lindblad  (GKSL) master equation~\cite{Manzano_2020AIPA},
\begin{align}
    \dot{\rho} = -i[\mathcal{H},\rho] + \sum_{n} (L_{n}\rho L_{n}^{\dag} - \frac{1}{2}\{L_{n}^{\dag}L_{n},\rho\}),
\end{align}
for cases C and F, where $L_{n}$ is the Lindblad operators.

\begin{figure}
    \centering
    \includegraphics[width = 8cm]{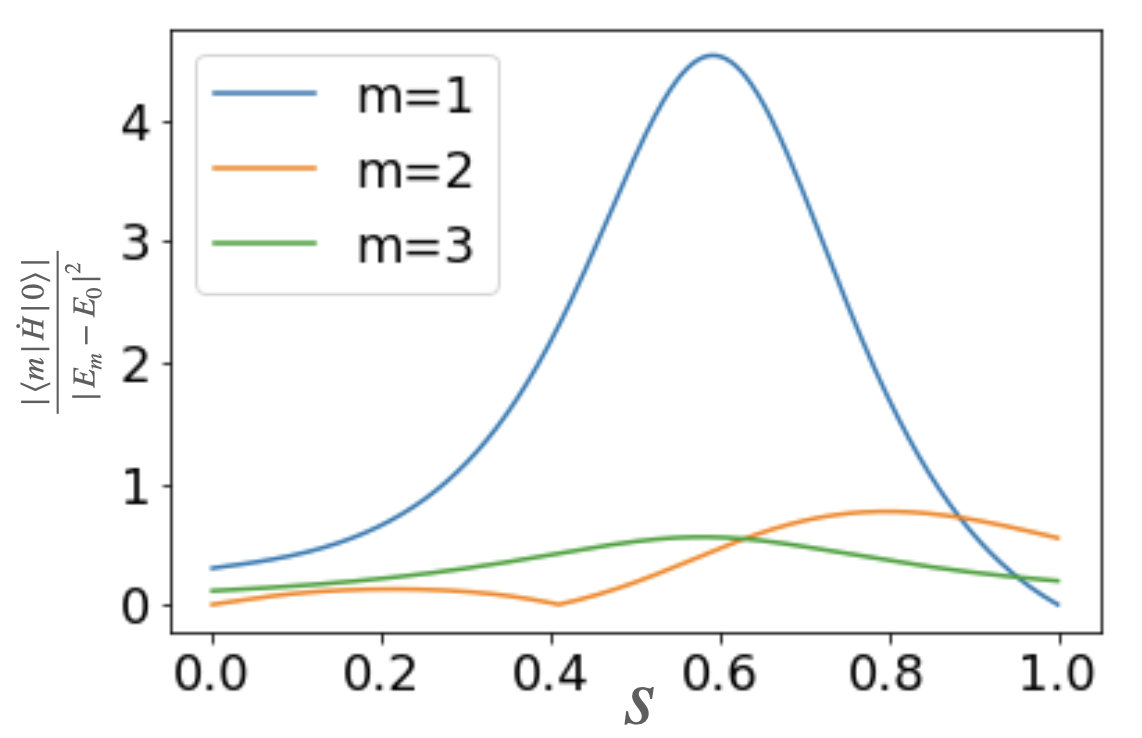}
    \caption{The actual adiabatic conditions~\eqref{eq:adiabatic_criterion} of our simulated two-qubit cases. $m=1$ is the largest at almost all $s$.}
    \label{fig:adcon}
\end{figure}

In the fifth step, we assume that an ideal projective measurement into the state $\ket{m(t=0)}$ can be performed. It is worth mentioning that, in the actual experiment, this projective measurement corresponds to $\sigma_{x}$ on all qubits. By using a post processing with a classical computer, we can obtain the projection probability for not only $m=1$ but also all $m$. However, the non-adiabatic transitions between the ground state and the first excited state is considered as the most relevant part. Actually, as long as $|\braket{1|\dot{H}|0}|$ is similar to or larger than $|\braket{m|\dot{H}|0}|$ for $m\geq 2$, the non-adiabatic transitions between the ground state and the first excited state is more relevant than the others, Thus, for the numerical simulations, we consider a case of $m=1$ in this paper. (See Fig.~\ref{fig:adcon})

\subsubsection{Construction of $\Omega_{\mathrm{exp}}(\omega)$}
\label{subsubsec:construction_Omegaomega}
In our method, we calculate the probability of a projection into the first excited state $p_{0,1}(\tau)$, and we use the power spectrum $P(\Omega)$ to determine the function $\Omega_{\mathrm{exp}}(\omega)$ as explained in the previous section.
In this case, we expect to observe a peak at $\Omega=\Omega_{\mathrm{ana}}(\omega)$ in the power spectrum. To determine the function $\Omega_{\mathrm{exp}}(\omega)$, we fix $\omega$ and maximize the height of the power spectrum by sweeping $\Omega$ so that we can determine the position of the resonance peak as follows:
\begin{align}
    \Omega_{\mathrm{exp}}(\omega) = \argmax_{\Omega} P(\Omega, \omega).\label{def:omega_est}
\end{align}
Finally, by sweeping $\omega$, we can obtain the function $\Omega_{\mathrm{exp}}(\omega)$.

When we sweep $\Omega$, it is crucial to choose an appropriate range.
First, we explore the frequency range $\Omega>0$. As indicated by Eq.~\eqref{eq:ft-ideal-output}, three peaks emerge. However, to evaluate the adiabatic condition, our focus lies solely on the positive-frequency peak, because the negative frequency peak contains the same information as its positive counterpart, while the zero frequency peak lacks relevant information.
Also, we should consider only the frequency range of $\Omega \ll \omega $ because we use RWA to derive the analytical formula of Eq.~\eqref{eq:general-Rabi}, which is valid only for $\Omega \ll \omega$.

Even if we restrict the frequency range, we may not find a correct peak for several reasons. We discuss the case in which such a problem occurs, and we present a possible solution to overcome such a problem at least for some cases.

\subsection{Single-qubit cases (A, B, and C)}
\label{subsec:single-qubit}

\begin{figure}[h]
    \begin{center}
    \begin{tabular}{c}
    \includegraphics[height = 4.5cm]{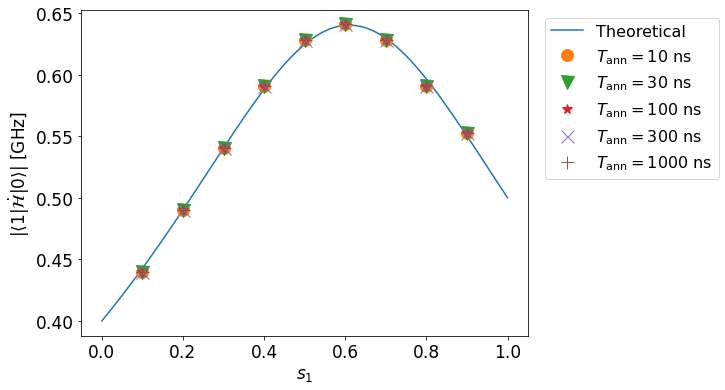}\\
    \includegraphics[height = 4.8cm]{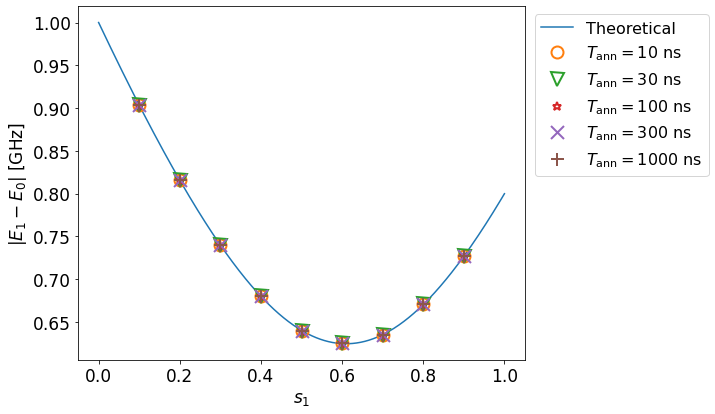}
    \end{tabular}
    \caption{(Top) Estimation of the transition matrix element in case A (single qubit, complete adiabaticity, and no decoherence).  (Bottom) Estimation of the energy gap in case A. The solid lines represent the solution obtained by diagonalization of the Hamiltonian, and the dots represent the estimated values obtained from our method by numerical simulation.}
    \label{fig:1qb-diag-noiseless}
    \end{center}
\end{figure}

We examine the single-qubit cases (A, B, and C). For these cases, the driver Hamiltonian $\mathcal{H}_{\mathrm{D}}$ and the problem Hamiltonian $\mathcal{H}_{\mathrm{P}}$ are given by
\begin{align}
    \mathcal{H}_{\mathrm{D}} = \frac{\omega_{1}}{2}\sigma_{x},\quad \mathcal{H}_{\mathrm{P}} =g\sigma_{z},\label{eq:onequbit-Hamiltonian}
\end{align}
respectively. In our simulation, we fixed $\omega_{1}=1\ \mathrm{GHz}$ and $g = 0.4\ \mathrm{GHz}$.

\subsubsection{Case A}
\label{subsubsec:caseA}

We set the parameters $T_\mathrm{ann}$ and $s_{1}$ as follows:
\begin{align}
    T_{\mathrm{ann}} &=10,~ 30,~ 100,~ 300,~ 1000\ \mathrm{ns},\nonumber\\
    s_{1} &= 0.1,~ 0.2,~ 0.3,~ 0.4,~ ..., 0.9.
\end{align}
As shown in Fig.~\ref{fig:1qb-diag-noiseless}, our estimated values (dots in the figure) are in good agreement with the theoretically expected values (lines in the figure). Indeed, the relative error in the estimation of the transition matrix element $|\braket{1|\dot{\mathcal{H}}|0}|$ (the energy gap $E_{1}-E_{0}$) is at most $0.99~\%$ ($0.071~\%$).

These errors are small compared to the resolution owing to the discretization performed while processing the data. The estimation error of the transition matrix element (energy gap) is $0.9$ ($0.1$) times smaller than the resolution. As shown in Fig.~\ref{fig:1qb-diag-noiseless}, we confirm that the adiabatic condition \eqref{eq:adiabatic_criterion} is reasonably satisfied.
\subsubsection{Case B}
\label{subsubsec:caseB}

\begin{figure}[h]
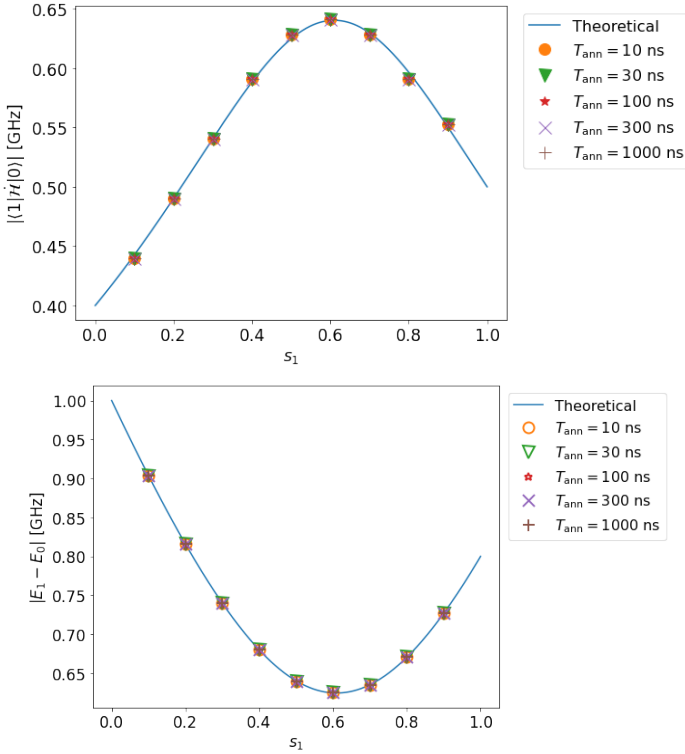

    \begin{tabular}{c}
    \includegraphics[height = 5cm]{1qb-ad-noiseless-matel-stype.png}\\
    \includegraphics[height = 5cm]{1qb-ad-noiseless-e01.png}
    \end{tabular}
    \caption{Top (bottom): estimated value of the transition matrix element (energy gap) in case B (single qubit, incomplete adiabaticity, and no decoherence).
    For the solid lines and dots, we use the same notation as that in Fig.~\ref{fig:1qb-diag-noiseless}.
    }
    \label{fig:1qb-ad-noiseless}
\end{figure}

\begin{figure}[h]
    \begin{tabular}{c}
    \includegraphics[height = 5cm]{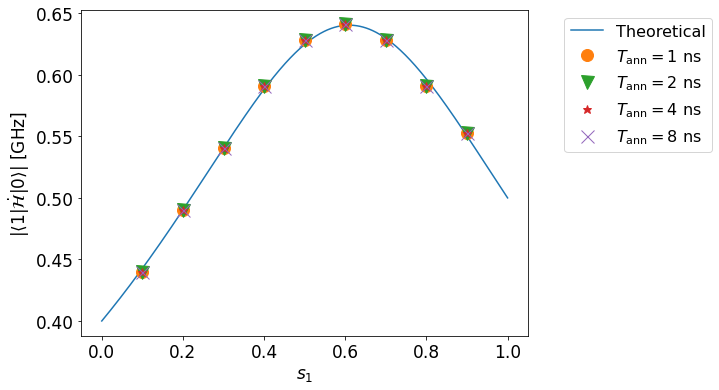}\\
    \includegraphics[height = 5cm]{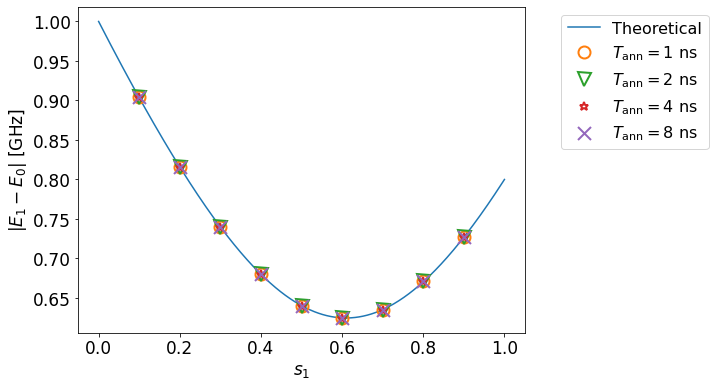}
    \end{tabular}
    \caption{Top (bottom): estimated value of the  transition matrix element (energy gap) in case B (single qubit, incomplete adiabaticity, and no decoherence) with a shorter annealing time such as
    $T_{\rm ann} = 1, 2, 4, 8$ $\mathrm{ns}$.
     For the solid lines and dots, we use the same notation as that in Fig.~\ref{fig:1qb-diag-noiseless}.}
    \label{fig:1qb-ad-noiseless-add}
\end{figure}

Next, the effect of non-adiabatic transitions in steps 2 and 4 is studied for case B.
Similar to case A, we can accurately measure both the transition matrix element $|\braket{1|\dot{\mathcal{H}}|0}|$ and the energy gap $(E_{1}-E_{0})$ for case B, and (see Fig.~\ref{fig:1qb-ad-noiseless}) the relative error of the transition matrix element (energy gap) is at most $2.2~\%$ ($0.7~\%$) and $0.77$ ($0.93$) relative to the resolution.

For the single-qubit case, our scheme is robust against the non-adiabatic transitions. Actually, we consider cases with $T_\mathrm{ann}=1, 2, 4$, and $8$ $\mathrm{ns}$ (see Fig.~\ref{fig:1qb-ad-noiseless-add}), and these results show that a shorter annealing time does not impair the performance of our methods.

We show that, as long as RWA is valid, the power spectrum contains a peak corresponding to a frequency of $\Omega(\omega)$ (see Appendix \ref{subsec:apb2-supposcase}). Thus, we can accurately estimate the transition matrix element and energy gap using Eqs.~\eqref{eq:numerator_est} and \eqref{eq:delta_est} for the single-qubit case without decoherence.

\subsubsection{Case C}
\label{subsubsec:caseC}

\begin{figure}[h]
    \begin{tabular}{c}
    \includegraphics[height = 4.8cm]{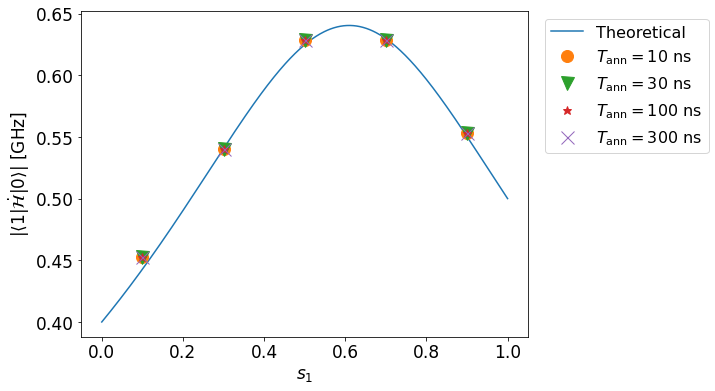}\\
    \includegraphics[height = 4.5cm]{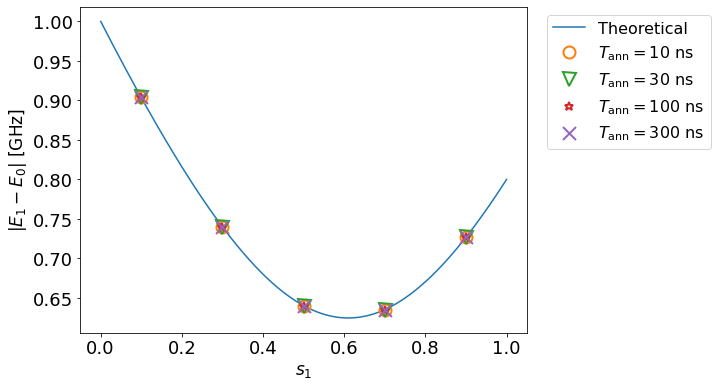}
    \end{tabular}
    \caption{
    Top (bottom): estimated value of the transition matrix element (energy gap) in case C (single qubit, incomplete adiabaticity, and decoherence).  For the solid lines and dots, we use the same notation as that in Fig.~\ref{fig:1qb-diag-noiseless}.
    }
    \label{fig:1qb-ad-noiz}
\end{figure}

In case C,
to consider decoherence, we employ the GKSL master equation, and we select the Lindblad operator as
\begin{align}
    L=\sqrt{\kappa} \sigma_{z},
\end{align}
where $\kappa$ denotes the decay rate. We fix $\kappa =2.5\times 10^{-3}$ $\mathrm{ns^{-1}}$, which is a typical value for a superconducting flux qubit \cite{yoshihara2006decoherence}.

The results are shown in Fig.~\ref{fig:1qb-ad-noiz}.
The relative error of the transition matrix element $|\braket{1|\dot{\mathcal{H}}|0}|$ (the energy gap $\Delta$) is at most $2.1~\%$ ($0.05~\%$), which is $0.77$ ($0.023$) times smaller than the resolution.

These errors are as small as those in cases A and B, indicating the robustness of our method against decoherence. This resilience stems from the fact that decoherence primarily impacts the width rather than the position of the peaks in the power spectrum. Consequently, accurate estimation of the transition matrix element and energy gap remains achievable even in the presence of weak decoherence.

\subsection{Two-qubit cases (D, E, and F)}
\label{subsec:Two-qubit}

\begin{figure}
    \begin{tabular}{c}
    \includegraphics[height = 5cm]{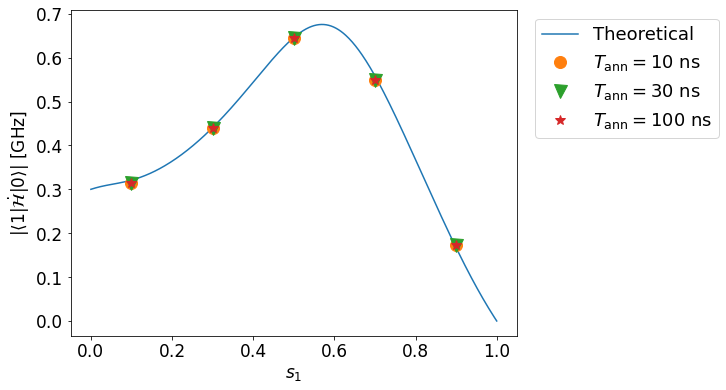}\\
    \includegraphics[height = 5cm]{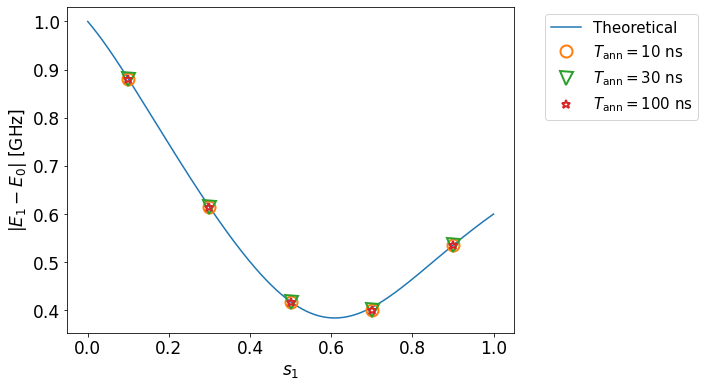}
    \end{tabular}
    \caption{Top (bottom): estimated value of the  transition matrix element (energy gap) in case D (two qubits, complete adiabaticity, and no decoherence). Even when two qubits are used, we can estimate both the transition matrix element and the energy gap with high accuracy.  For the solid lines and dots, we use the same notation as that in Fig. \ref{fig:1qb-diag-noiseless}.}
    \label{fig:2qb-diag-noiseless}
\end{figure}

In the two-qubit cases, the problem and driver Hamiltonians are given by
\begin{align}
    \mathcal{H}_{\mathrm{D}} &= \frac{\omega_{1}}{2}\sigma_{x}\otimes 1 + \frac{\omega_{2}}{2}1\otimes \sigma_{x},\nonumber\\
    \mathcal{H}_{\mathrm{P}} &=g_{1}\sigma_{z}\otimes\sigma_{z} + g_{2}\sigma_{z}\otimes 1
    + g_{3}1\otimes \sigma_{z},
    \label{eq:twoqubit-Hamiltonian}
\end{align}
respectively. Here, we set $\omega_{1} = 1.0\ \mathrm{GHz}$, $\omega_{2} = 1.1\ \mathrm{GHz}$, $g_{1} = 0.5\ \mathrm{GHz}$,  $g_{2}=0.3\ \mathrm{GHz}$, and $g_{3}=0$.

For these cases, we select the parameters $T_\mathrm{ann}$ and $s_{1}$ as follows.

\begin{align}
    T_{\mathrm{ann}} &=10,~ 30,~ 100\ \mathrm{ns,}\nonumber\\
    s_{1} &= 0.1,~ 0.3,~ 0.5,~ 0.7,~ 0.9.
\end{align}

\subsubsection{Case D}
\label{subsubsec:case_D}

In this case, we can accurately measure the transition matrix element $|\braket{1|\dot{\mathcal{H}}|0}|$ and the energy gap $(E_{1}-E_{0})$ as shown in Fig.~\ref{fig:2qb-diag-noiseless}. The relative error of the transition matrix element (energy gap) is at most $3.5~\%$ ($0.04~\%$), which is $0.55$ ($0.99$) times smaller than the resolution. Despite not satisfying condition III for considering two qubits in this case, the dynamics can be effectively confined within a two-level system, ensuring the accuracy of our method, especially when the Rabi frequency is low.

\subsubsection{Case E}
\label{subsubsec:case_E}

\begin{figure}[ht]
    \begin{tabular}{c}
    \includegraphics[height = 4.8cm]{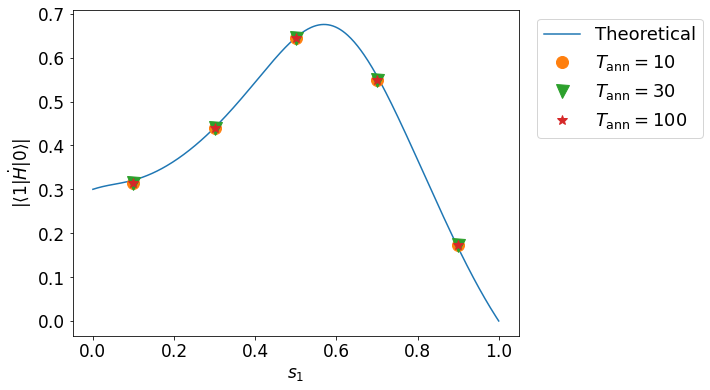}\\
    \includegraphics[height = 4.8cm]{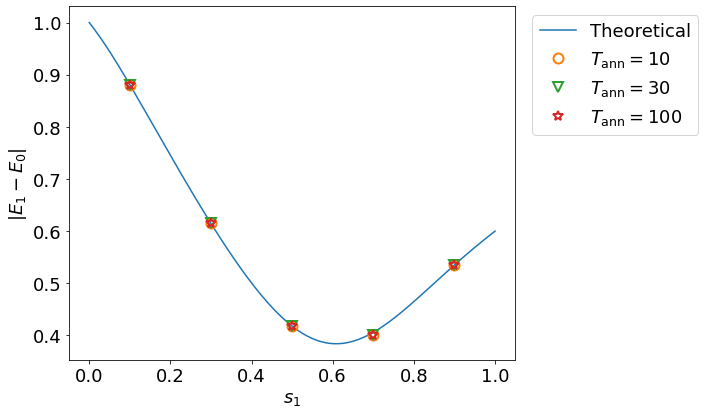}
    \end{tabular}
    \caption{Top (bottom): estimated value of the transition matrix element (energy gap) in case E (two qubits, incomplete adiabaticity, and no decoherence). 
     For the solid lines and dots, we use the same notation as that in Fig.~\ref{fig:1qb-diag-noiseless}.
    }
    \label{fig:2qb-ad-noiseless}
\end{figure}

In case E, the relative error of the transition matrix element (energy gap) is at most $3.5~\%$ ($1.2~\%$), which is $0.55$ ($0.99$) times smaller than the resolution, as shown in Fig.~\ref{fig:2qb-ad-noiseless}.

In the case of weak non-adiabatic transitions, it is possible to estimate both the transition matrix element and the energy gap with high accuracy even for the two-qubit case. Meanwhile, as described in detail in Appendix \ref{sec:nonadiabatic}, in the case of strong non-adiabatic transitions, the power spectrum contains peaks other than the one that we want to use in our estimation.
We discuss a possible solution for this problem in Appendix~\ref{sec:nonadiabatic}.

\subsubsection{Case F}
\label{subsubsec:case_F}

\begin{figure}[h!]
    \begin{tabular}{c}
    \includegraphics[height = 5cm]{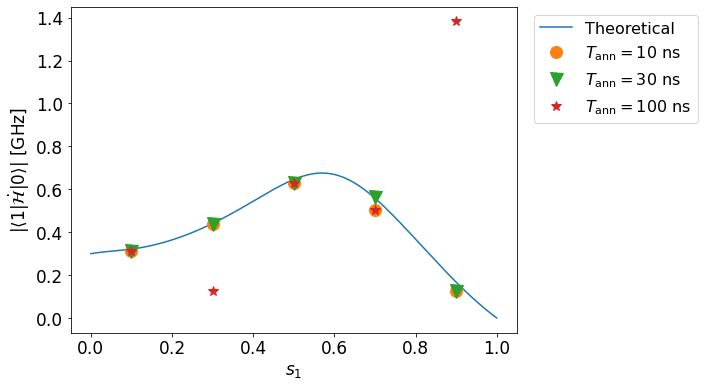}\\
    \includegraphics[height = 5cm]{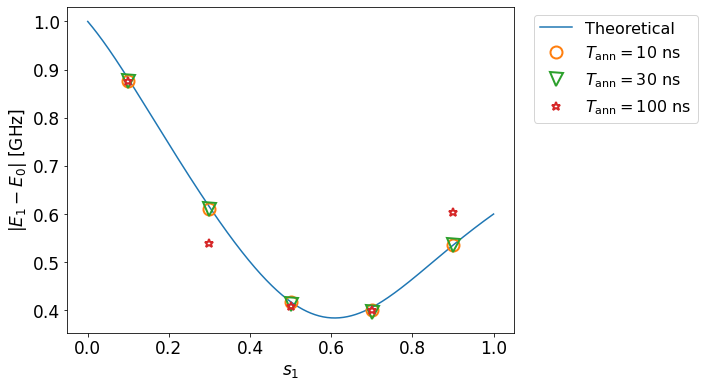}
    \end{tabular}
    \caption{Top (bottom): estimated value of the transition matrix element (energy gap) in case F (two qubits, incomplete adiabaticity, and decoherence).
    In this case, we have significant estimation errors for a few points.
     For the solid lines and dots, we use the same notation as that in Fig.~\ref{fig:1qb-diag-noiseless}.}
    \label{fig:2qb-ad-noiz}
\end{figure}

In case F, we select the Lindblad operator as follows.
\begin{align}
    L_{1} =\sqrt{\kappa}\sigma_{z}\otimes 1,\qquad L_{2} = \sqrt{\kappa} 1\otimes\sigma_{z},
\end{align}
Here, $\kappa$ denotes the decay rate. For the numerical simulations, we chose $\kappa =2.5\times 10^{-3}$ $\mathrm{ns}^{-1}$.

\begin{figure}[h!]
    \begin{tabular}{c}
    \includegraphics[width = 8cm]{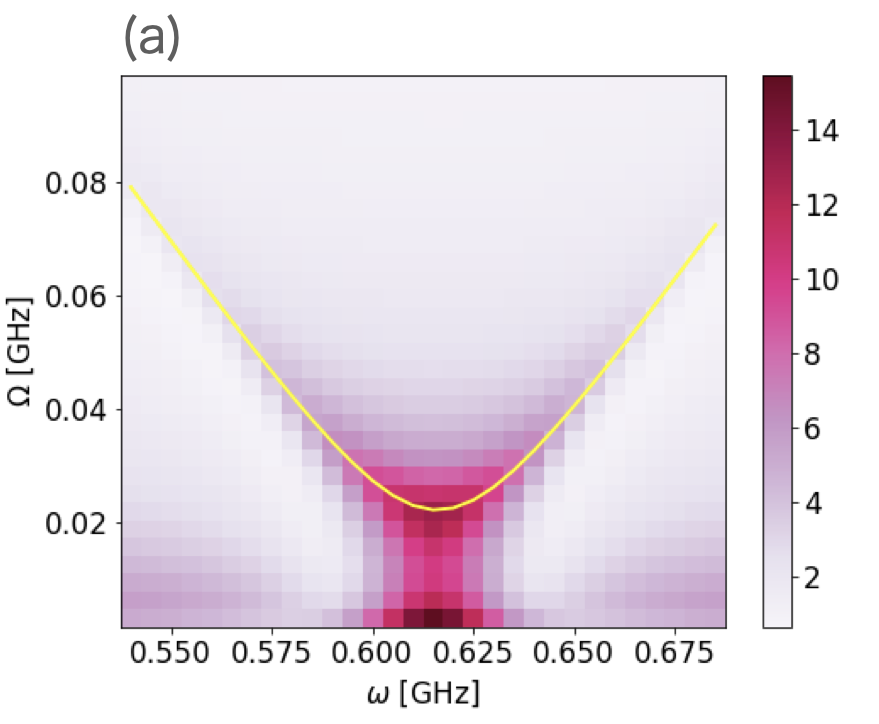}\\
    \includegraphics[width = 8cm]{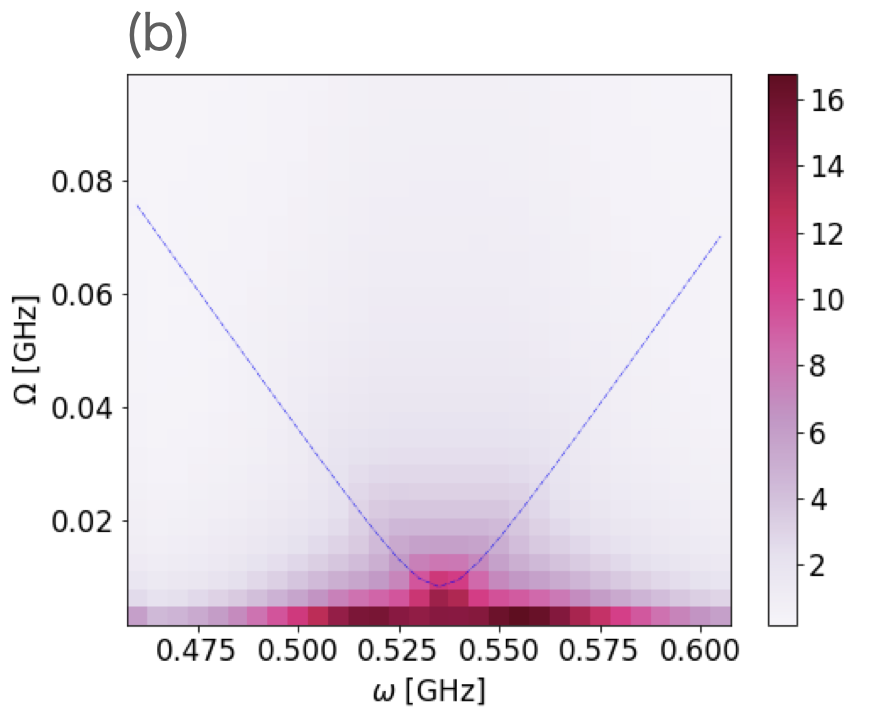}
    \end{tabular}
    \caption{Plot of the power spectrum $P(\omega,\Omega)$ for $T_\mathrm{ann}=100$ for case F (two qubits, incomplete adiabaticity, and decoherence).
    The horizontal axis represents the angular frequency of the driving field and the vertical axis represents the Fourier frequency.
    (a) We use $s_{1}=0.3$.
    The yellow line is obtained by fitting Eq.~\eqref{eq:hyperbolic-curve} to the plot.
    (b) We use $s_{1}=0.9$.
    The dotted line represents the exact value obtained by diagonalization. }
    \label{fig:2qb-ad-noiz-sps03}
\end{figure}

We plot the estimated transition matrix element and energy gap against $s_{1}$, and we demonstrate that our method is accurate except for two points, $s_{1}=0.3$ and $s_{1}=0.9$ for $T_\mathrm{ann}=100\ \mathrm{ns}$, as shown in Fig.~\ref{fig:2qb-ad-noiz}. In the former case, as shown in the power spectrum (see Fig.~\ref{fig:2qb-ad-noiz-sps03} (a)), where $\omega$ is smaller than $0.575$ or larger than $0.65$, a low-frequency ($\Omega < 0.02$) peak exists, and the height of this peak is greater than that of the target peak at the same $\omega$.

As shown in Eq.~\eqref{eq:ft-ideal-output}, strictly speaking, a peak around $\Omega \simeq 0$ should exist in the spectrum, and this peak has a finite width owing to decoherence so that we can observe this in case F. Therefore, if we naively adopt our method described in Eq.~\eqref{def:omega_est}, we generate an inappropriate $\Omega_{\rm exp}(\omega)$ and obtain incorrect estimated values of the transition matrix element and energy gap.

To identify the target peak in the presence of decoherence and non-adiabatic conditions, we employ a modified approach outlined as follows. Initially, we assess the value of $\Omega$ not only for the highest peak but also for the second- and third-highest peaks at each $\omega$. These pairs of values $(\omega, \Omega)$ then constitute candidates for the data in the estimated function $\Omega_{\rm exp}(\omega)$. Subsequently, we attempt to fit the data using the analytical formula presented in Eq.~\eqref{eq:hyperbolic-curve}. In the third step, we eliminate data that cannot be adequately fitted by the analytical formula. Finally, we designate data successfully fitting the analytical formula as the target peaks.

In the former case ($s_{1}=0.3$), after using this modified method, the relative error of the transition matrix element (energy gap) is $1.0\%$ $(0.2\%)$, and the ratio to the resolution is $0.07$ $(0.26)$.
Thus, our modified method is effective for this case, as shown in Fig.~\ref{fig:2qb-ad-noiz-sps03} (a).

However, in the latter case ($s_{1}=0.9$), the decoherence is so strong that the target peak nearly disappears, and we cannot identify the target peak anymore, as shown in Fig.~\ref{fig:2qb-ad-noiz-sps03} (b).

\section{Conclusions and discussion}
\label{sec:conclusion_and_discussion}
We have proposed an experimental method to assess adiabaticity in QA by evaluating  adiabatic conditions. Our approach uses an oscillating field to induce Rabi oscillations, providing insights into the energy gap and the transition matrix element of the time derivative of the Hamiltonian. To validate our method, we performed numerical simulations, considering non-adiabatic transitions and decoherence effects. The results confirm the robustness of our method against these experimentally inevitable problems.

Our method is valuable for determining an optimal annealing schedule and Hamiltonian form to improve the performance of QA. When a phase transition takes place, the performance of QA is degraded. Some methods have been proposed to address this issue in specific cases \cite{Jorg_2010EPL, Seki_2019_JPSJ, Watabe_2020}.
To apply these methods, we need to change the annealing schedule and form of the Hamiltonian. However, a potential problem is that we cannot easily optimize the annealing scheduling and form of the Hamiltonian for general problems if we do not know whether the adiabatic conditions are satisfied. Meanwhile, by using our methods to evaluate the adiabaticity of the dynamics, we can select an appropriate annealing schedule and form of the Hamiltonian when we try to solve practical optimization problems using QA.

Also, we discuss possible experimental implementations of our proposal. 
In the currently available D-wave quantum annealing machine, it is not possible to perform our proposal because we cannot perform microwave pulses to induce the Rabi oscillation in the annealer. However, the recently proposed method, called the spin-lock quantum annealing \cite{Matsuzaki_2020JJAP, Imoto_seki_2022}, is compatible with the requirement of our method.
In the spin-lock quantum annealing, we use superconducting qubits for a gate type quantum computer, and there is an experimental demonstration of the spin-lock with superconducting qubits by using microwave pulses~\cite{Abdurakhimov_2020PhysRevB}. 
Let us also examine the validity of the parameters used in our numerical simulations.
During the spin lock, the system is in the rotating frame and so we can set $\omega_{1} = 1.0~\mathrm{GHz}$ and $\omega_{2} = 1.1~\mathrm{GHz}$ as the detuning between the microwave frequency and qubit resonance.
The reported coherence time of the superconducting qubit~\cite{Bylander_2011natphys} is much longer than $1/\kappa = 400~\mathrm{ns}$ which is used in our simulations.
Also, it is possible to realize a coupling strength of $g_{1} = 0.5~\mathrm{GHz}$ and $g_{2}=0.3~\mathrm{GHz}$
by using inductive coupling between the superconducting qubits~\cite{majer2005spectroscopy}. 
We set the Rabi frequency $\lambda$ to be around tens of $\mathrm{MHz}$, which is also available in the current experiment~\cite{Yoshihara_2014PRB}.

It is important to note that our method remains applicable even if the number of qubits increases.
Even in such scenarios, there 
must be finite energy gaps between the eigenenergies.
Our approach is designed to effectively operate in such situations (as long as the energy gap is finite) where long-lived qubits are available for quantum annealing.

In QA, the energy gap could approach to zero as we increase the size of the system. In such a case, we will find a highly degenerate spectrum.
In our method, we sweep $s_1$ and investigate the adiabatic condition for several values of $s_1$.
Since we sweep the value of $s_1$ from $0$ to another value in our method, we can study the adiabatic condition during QA before the energy gap closes. In this case, we will recognize that the energy gap becomes smaller as we increase $s_1$, and this lets us know the existence of the energy gap closing.
We could adopt several strategies to enlarge the energy gap in this case. For example, twisted field~\cite{Kadowaki2021_ptrsa, Imoto2022_NJP}, counterdiabatic term~\cite{Hartmann2019_NJP, Hayasaka2022_arxiv}, nonstoquastic Hamiltonian~\cite{Seki2012_PRE,Hormozi2016_PRB, Susa2022_arXiv}, inhomogeneous driving magnetic field~\cite{Susa2019_JPSJ} are helpful for such a purpose. 
It is considered as difficult to find such additional fields to resolve the problem of the energy gap closing.
Importantly, our method is useful to find such additional fields for the following reasons.
 When we introduce the additional fields, the adiabatic conditions will be changed,
  and we can experimentally measure the adiabatic conditions from the spectrum. So, if introducing the additional fields improves the adiabatic conditions, we can detect it. In principle, by measuring the adiabatic conditions, we can variationally change the strength and the direction of the additional fields to optimize the adiabatic conditions. The detailed research about this is left for future work. In the revised manuscript, we explained this point.

Finally, we comment on the adiabatic condition itself. In general, it has not been proved that the condition~\eqref{eq:adiabatic_criterion} is 
sufficient to achieve the adiabaticity.
In addition, more sophisticated criteria have been proposed~\cite{Jansen_2007JMP},
and it was shown that higher order derivative of the annealing Hamiltonian could affect the adiabaticity. 
In our numerical examples, we show that
the condition~\eqref{eq:adiabatic_criterion} actually provides an upper bound of the population of the excited state due to the non-adibatic transitions in Appendix~\ref{sec:valid}. 
However, if we need to know the information of the higher order derivative of the Hamiltonian, we can use a modified version of our method. For example, if we are interested in the value of $|\braket{m|\ddot{H}|0}|$, we can replace
$\dot{H}(s_{1})$ in Eq.~\eqref{def:external_Hamiltonian} with $\ddot{H}(s_{1})$.
We leave a detailed study of this for future work. Secondly, although the conventional adiabatic condition in Eq.~\eqref{eq:adiabatic_criterion} is derived from the unitary evolution, it is possible to generalize the adiabatic theorem to open quantum systems~\cite{Venuti_2016PRA}.
The aim of our method is to know the value of Eq.~\eqref{eq:adiabatic_criterion}, which is different from the adiabatic condition in the open quantum systems. 
We also leave the extension of our method to the adiabatic condition in open quantum systems for future work.

\begin{acknowledgments}
We are grateful to Takashi Imoto, Hideaki Okane, Hiroshi Hayasaka, and Tadashi Kadowaki for their insightful comments.

This work was supported by the Leading Initiative for Excellent Young Researchers, MEXT, Japan, and JST Presto
(Grant No.~JPMJPR1919), Japan. This paper is partly
based on the results obtained from a project, JPNP16007,
commissioned by the New Energy and Industrial Technology Development Organization (NEDO), Japan.

We thank the developers of QuTiP \cite{Qutip}, which was used for our numerical simulations.
\end{acknowledgments}

\appendix
\section{Adiabatic theorem and adiabatic condition}
\label{sec:adiaba-condi}

In this section, we review the adiabatic theorem. We consider a time-dependent Hamiltonian $\mathcal{H}(s)$.
For each time $s$, we denote the eigenstates (called instantaneous eigenstates) obtained by diagonalizing the Hamiltonian $\mathcal{H}(s)$ as $\ket{n(s)}$ and the eigenvalues (called instantaneous energy) as $E_n(s)$.
\begin{align}
    \mathcal{H}(s)\ket{n(s)}=E_{n}(s)\ket{n(s)}\label{eq:hamil_eigeq}
\end{align}
For any state $\ket{\psi(s)}$, at each time $s$, the state can be expanded using the instantaneous eigenstates $\ket{n(s)}$ as follows:
\begin{align}
    \ket{\psi(s)}=\sum_{n}c_{n}(s)e^{-isT_{\rm ann}\bar{E}_{n}(s)}\ket{n(s)},
\end{align}
where $\bar{E}_{n}(s)$ is defined by
\begin{align}
    \bar{E}_{n}(s) = \frac{1}{s}\int_{0}^{s} d\sigma E_{n}(\sigma).
\end{align}
Since this $\ket{\psi(s)}$ is a solution of the Schroedinger equation, the state satisfies
\begin{align}
    i\frac{d}{ds}\ket{\psi(s)}&=i\sum_{n}\frac{d}{ds}(c_{n}(t)e^{-isT_{\rm ann}\bar{E}_{n}(s)}\ket{n(s)})\nonumber\\
    &=i\sum_{n}\dot{c}_{n}(s)e^{-isT_{\rm ann}\bar{E}_{n}(s)}\ket{n(s)} \nonumber\\
    &\quad + (-i\frac{d}{ds}(sT_{\rm ann}\bar{E}_{n}(s)))c_{n}(s)e^{-isT_{\rm ann}\bar{E}_{n}(s)}\ket{n(s)}\nonumber \\
    &\quad+c_{n}(s)e^{-isT_{\rm ann}\bar{E}_{n}(s)}\ket{\dot{n}(s)}\nonumber\\
    &=\sum_{n}c_{n}(s)T_{\rm ann}E_{n}(s)e^{-isT_{\rm ann}\bar{E}_{n}(s)}\ket{n(s)}.\label{eq:schr-ad}
\end{align}
Combining Eq.~\eqref{eq:schr-ad} and the orthonormality of the eigenstates, we obtain
\begin{align}
    &i\dot{c}_{n}(s)e^{-isT_{\rm ann}\bar{E}_{n}(s)}+i\sum_{m}c_{m}(s)e^{-isT_{\rm ann}\bar{E}_{m}(s)}\braket{n(s)|\dot{m}(s)}\nonumber \\
    &=0.\label{eq:pre-result}
\end{align}
Now, we differentiate both sides of Eq.~\eqref{eq:hamil_eigeq} at time $s$:
\begin{align}
    \dot{\mathcal{H}}(s)\ket{n(s)}+\mathcal{H}(s)\ket{\dot{n}(s)}=\dot{E}_{n}(s)\ket{n(s)}+E_{n}(s)\ket{\dot{n}(s)}.
\end{align}
By taking the inner product with $\ket{m(s)}$ again, Eq.~\eqref{eq:pre-result} becomes
\begin{align}
    &\dot{c}_{n}(s)+\braket{n(s)|\dot{n}(s)}c_{n}(s)\nonumber\\
    &=\sum_{m \neq n}\frac{\braket{n(s)|\dot{\mathcal{H}}|m(s)}}{E_{n}(s)-E_{m}(s)}c_{m}(s)e^{isT_{\rm ann}(\bar{E}_{n}(s)-\bar{E}_{m}(s))}.\label{eq:difeqofct}
\end{align}
When we ignore the right-hand side of Eq.~\eqref{eq:difeqofct}, we can get the adiabatic theorem. Indeed, it is clear that no transitions between the energy levels occur in this time evolution, because the differential equation contains only one variable $c_{n}(t)$.

To obtain the final state after the time evolution, we integrate
both sides of Eq.~\eqref{eq:difeqofct}, and we obtain
\begin{align}
    c_{n}&(s)-c_{n}(0)=-\int_{0}^{s}\braket{n(\sigma)|\dot{n}(\sigma)}c_{n}(\sigma) d\sigma\nonumber\\
   &+\int_{0}^{s}\sum_{m \neq n}\frac{\braket{n(\sigma)|\dot{\mathcal{H}}|m(\sigma)}}{E_{n}(\sigma)-E_{m}(\sigma)}c_{m}(\sigma)e^{iT_{\rm ann}\sigma(\bar{E}_{n}(\sigma)-\bar{E}_{m}(\sigma))} d\sigma.\label{eq:integral_equation}
\end{align}
If necessary, we perform a transformation of the basis $\ket{\tilde{n}(s)} =e^{i\theta(s)}\ket{n(s)}$, and $\braket{n(s)|\dot{n}(s)}$ can be zero. Thus, Eq.~\eqref{eq:integral_equation} becomes
\begin{align}
    &\quad c_{n}(s)-c_{n}(0)\nonumber\\
    &=\int_{0}^{s}\sum_{m \neq n}\frac{\braket{n(\sigma)|\dot{\mathcal{H}}|m(\sigma)}}{E_{n}(\sigma)-E_{m}(\sigma)}c_{m}(\sigma)e^{iT_{\rm ann}\sigma(\bar{E}_{n}(\sigma)-\bar{E}_{m}(\sigma))} d\sigma.\label{eq:integral_equation2}
\end{align}
By recursive use of Eq.~\eqref{eq:integral_equation2}, we obtain a form of $c_{n}(s)$ as an infinite series. If we use the first-order perturbation, we obtain
\begin{align}
    &c_{n}(s)\nonumber\\
    &\simeq c_{n}(0)+\sum_{m \neq n} c_{m}(0)\int_{0}^{s}\frac{\braket{n(\sigma)|\dot{\mathcal{H}}|m(\sigma)}}{E_{n}(\sigma)-E_{m}(\sigma)}e^{iT_{\rm ann}\sigma(\bar{E}_{n}(\sigma)-\bar{E}_{m}(\sigma))} d\sigma\nonumber\\
    &=c_{n}(0) -i\sum_{m\neq n}c_{m}(0)\left(A_{mn}(s)-A_{mn}(0) + B_{mn}(s)\right),\label{eq:integral_equation3}
\end{align}
where
\begin{align}
A_{mn}(s) &=\frac{\braket{n(s)|\dot{\mathcal{H}}|m(s)}}{T_{\rm ann}(E_{n}(s)-E_{m}(s))^{2}}e^{isT_{\rm ann}(\bar{E}_{n}(s)-\bar{E}_{m}(s))}, \label{eq:anm}\\
B_{mn}(s) &= \int_{0}^{s}\frac{e^{iT_{\rm ann}\sigma(\bar{E}_{n}(\sigma)-\bar{E}_{m}(\sigma))}}{T_{\rm ann}(E_{n}(\sigma)-E_{m}(\sigma))}\frac{d}{d\sigma}\left[\frac{\braket{n(\sigma)|\dot{\mathcal{H}}|m(\sigma)}}{E_{n}(\sigma)-E_{m}(\sigma)}\right]d\sigma.\label{eq:bnm}
\end{align}
Here, we use
\begin{align}
    \frac{d}{d\sigma}\left[T_{\rm ann}\sigma(\bar{E}_{n}(\sigma)-\bar{E}_{m}(\sigma))\right]=T_{\rm ann}(E_{n}(\sigma)-E_{m}(\sigma)).
\end{align}
When Eq.~\eqref{eq:adiabatic_criterion} is satisfied, $A_{mn}(s)$ is negligible. In general, if $B_{mn}(s)$ is nonzero, $c_{n}(s)$ is different from $c_{n}(0)$ in the first-order perturbation so that the adiabaticity is not always guaranteed by Eq.~\eqref{eq:adiabatic_criterion}~\cite{Amin_PRL2009, Dodin_Brumer_PRXQ2021}. However, if $E_{n}$ and $E_{m}$ are time-independent, $B_{mn}(s)$
corresponds to a Fourier transformation; hence, $B_{mn}(s)$ can be ignored
unless the integrating function includes a component whose angular frequency corresponds to the (average) energy gap $\bar{E}_{n}-\bar{E}_{m}$.
Except for the special non-negligible $B_{mn}$ case, the condition \eqref{eq:adiabatic_criterion} makes $c_{n}(s) = c_{n}(0)$, and it shows the statement of the adiabatic theorem.

Usually, when we consider QA, the initial states are designated as
\begin{align}
    c_{0} =1,\ c_{n}=0\ (n\neq 0),
\end{align}
so that we can finally obtain
\begin{align}
    c_{0}(s) = 1,\ c_{n}(s) = -i(A_{0n}(s)-A_{0n}(0)). \label{eq:approx_wavf}
\end{align}
Furthermore, $c_{n}(s)\simeq 0$ if the change in the Hamiltonian is sufficiently slow.

\section{Rabi oscillation}
\label{sec:Rabi-o}
We focused on the characteristics of the dynamics of the system in our proposal.
As explained in the main text, our scheme for a single qubit
is equivalent to the conventional Rabi oscillation as long as the dynamics at the second and forth steps are adiabatic. 
Meanwhile, if non-adiabatic transitions occur at the steps, the observed dynamics in our scheme deviates from the conventional Rabi oscillation.

\subsection{Conventional Rabi oscillation}
\label{ubsec:usual-rabio}
Based on the discussion in the main text,
we consider a Rabi oscillation between the states $|k\rangle $ and $|l\rangle$
(for example, the Hamiltonian in Eq.~\eqref{eq:external-hamil} corresponds to a case with $k=0$ and $l=m$). We have
\begin{align}
    \mathcal{H}_{\mathrm{eff}} =(1-r)\frac{\Delta}{2}\sigma_{z}+\frac{\tilde{\lambda}}{2}\ket{k}\bra{l}+\frac{\tilde{\lambda}^{*}}{2}\ket{l}\bra{k}, \label{eq:qubit_rwahamil}
\end{align}
where $\sigma_{z} = \ket{l}\bra{l}-\ket{k}\bra{k}$, $\Delta = E_{l}-E_{k}$, and $\tilde{\lambda} = \lambda\braket{l|\dot{\mathcal{H}}_{\rm QA}|k}$.
This coincides with the conventional Hamiltonian to induce the Rabi oscillation, where $\tilde{\lambda}$ denotes the Rabi frequency and $(1-r)\Delta$ denotes the detuning.
For this Hamiltonian, the operation $e^{i\theta \sigma_{z}}$ is adopted for rotation about the $z$-axis by an appropriate angle $\theta$. Then, the Hamiltonian \eqref{eq:qubit_rwahamil} becomes
\begin{align}
    \mathcal{H}_{\mathrm{eff}} =(1-r)\frac{\Delta}{2}\sigma_{z}+\frac{|\tilde{\lambda}|}{2}\sigma_{x}, \label{eq:qubit_rwahamil_mod}
\end{align}
where $\sigma_{x} = \ket{k}\bra{l}+\ket{l}\bra{k}$. We can rewrite Eq.~\eqref{eq:qubit_rwahamil_mod} with a unitary operator $U_{\rm diag}$ as
\begin{align}
    \mathcal{H}_{\mathrm{eff}} = \frac{\tilde{\lambda}'}{2}U^{\dag}_{\rm diag}\sigma_{x}U_{\rm diag},
\end{align}
where $\tilde{\lambda}'$ satisfies
\begin{align}
    \frac{\tilde{\lambda}'}{2}=\sqrt{\left(\frac{|\tilde{\lambda}|}{2}\right)^{2}+\left((1-r)\frac{\Delta}{2}\right)^{2}}.\label{eq:Omega-Rabi-squbit}
\end{align}
By setting $r=\omega/\Delta$, we obtain $\tilde{\lambda}'=\Omega_{\mathrm{ana}}^{(k,l)}(\omega)$ in Eq. \eqref{eq:hyperbolic-curve}.

\begin{figure}[htbp]
    \centering
    \includegraphics[width = 8cm]{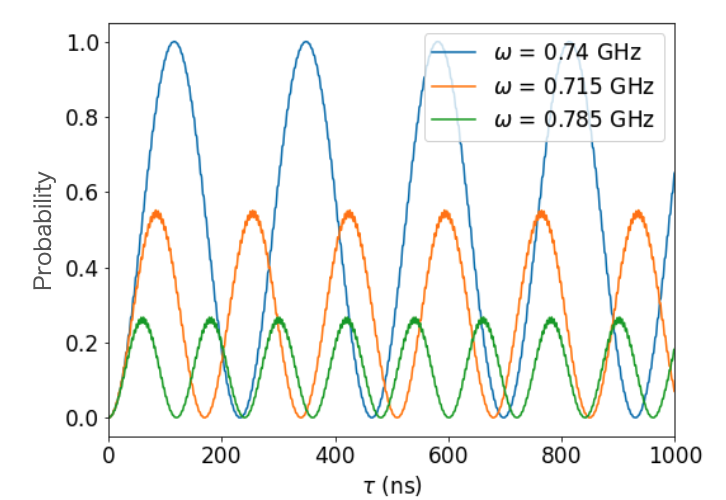}
    \caption{Plot of the Rabi oscillation.
    We use a single-qubit Hamiltonian \eqref{eq:onequbit-Hamiltonian},
    with $T_\mathrm{ann} = 30$ and $t_1 = 0.3\, T_\mathrm{ann}$. The oscillation period is the longest when the angular frequency $\omega$ of the external field coincides with the energy gap $\Delta = 0.74$, indicating that the oscillation period is shorter for small angular frequencies ($\omega = 0.715$) and large angular frequencies ($\omega = 0.785$). }
    \label{fig:Rabi-oscillation-onequbit}
\end{figure}

To observe the Rabi oscillations, we calculate the amplitude as follows:
\begin{align}
    \braket{l|e^{-it\mathcal{H}_{\mathrm{eff}}}|k}&=\braket{l|U^{\dag}_{\rm diag}e^{-it\frac{\tilde{\lambda}'}{2}\sigma_{x}}U_{\rm diag}|k},\nonumber\\
    &=-i\sin{\frac{\tilde{\lambda}'}{2} t}\braket{l|U^{\dag}_{\rm diag}\sigma_{x}U_{\rm diag}|k},
\end{align}
using the relations
\begin{align}
    &e^{i\theta\sigma_{x}}=\cos{\theta}+i\sin{\theta}\sigma_{x},\\
    &\braket{l|U^{\dag}_{\mathrm{diag}}U_{\mathrm{diag}}|k}=0.
\end{align}
Finally, we obtain
\begin{align}
    |\braket{l|e^{-it\mathcal{H}_{\mathrm{eff}}}|k}|^{2}=|\braket{l|U^{\dag}_{\rm diag}\sigma_{x}U_{\rm diag}|k}|^{2}\frac{1-\cos{\tilde{\lambda}' t}}{2}.\label{eq:rabi-oscil-calc-ideal}
\end{align}
Owing to the form of Eq.~\eqref{eq:Omega-Rabi-squbit}, the minimum value of $\tilde{\lambda}'$ is given by $s = 1$; then, the angular frequency of the Rabi oscillation $\tilde{\lambda}'$ is $|\tilde{\lambda}|=\lambda|\braket{l|\dot{\mathcal{H}}|k}|$. In this case, $U_{\rm diag} = I$ and the amplitude is maximized. Thus, the Rabi oscillation is a type of resonance phenomenon whose resonant angular frequency is the energy gap $\Delta$.

When the dynamics in the second and fourth steps is adiabatic, both the initial state and the final state are energy eigenstates. Meanwhile, Fig.~\ref{fig:Rabi-oscillation-onequbit} shows that the Rabi oscillations were obtained numerically using the Hamiltonian induced by Eq. \eqref{eq:onequbit-Hamiltonian} directly, without the approximation described in this section. It is generally a simple sinusoidal curve; however, there are slight fine oscillations when the details are observed.

\subsection{Dynamics with non-adiabatic conditions}
\label{subsec:apb2-supposcase}
We explain the dynamics of our system when the non-adiabatic transitions occur in the second and fourth steps. Here, for simplicity, we assume that RWA is valid. Owing to the non-adiabatic transitions, in general, our prepared state at time $t_{1}$ is not an energy eigenstate; it is a superposition state of the energy eigenstates. Similarly, the non-adiabatic transition also occurs from $t_{1}+\tau$ to $2t_{1}+\tau$. We will explain that these non-adiabatic transitions cause high-frequency oscillations in our scheme.

In step 3, we prepare a state $|\psi(0) \rangle $ and let this state evolve by the Hamiltonian for time $t$. The state we obtain at time $t$ denotes $|\psi(t) \rangle $. Combining the unitary evolution in step 4 with the projective measurement performed in step 5, this process can be considered as a projective measurement $|\phi (t)\rangle \langle \phi (t)|$ on the state $|\psi(t) \rangle $.

An overlap between the states is given by $\braket{\phi(t)|\psi(t)} = \braket{\tilde{\phi}(t)|\tilde{\psi}(t)}$, where we define
\begin{align}
    \bra{\tilde{\phi}(t)}=\bra{\phi(t)}e^{-irt\mathcal{H}_{\mathrm{QA}}}.
\end{align}
Using RWA, the final transition amplitude is calculated as
\begin{align}
    \braket{\tilde{\phi}(t)|\tilde{\psi}(t)} &= \braket{\tilde{\phi}(t)|e^{-it\mathcal{H}_{\mathrm{eff}}}|\tilde{\psi}(0)},\nonumber\\
    &= \braket{\phi(t)|e^{-irt\mathcal{H}_{\mathrm{QA}}}e^{-it\mathcal{H}_{\mathrm{eff}}}|\psi(0)},\label{eq:traamp}
\end{align}
where we use the effective Hamiltonian described in Eq.~\eqref{eq:external-hamil}.
As we prepare a superposition of different energy eigenstates by a non-adiabatic transition and then perform a projective measurement onto another superposition of the energy eigenstates, the difference between the energy eigenvalues affects the oscillation.  We will demonstrate this point below.

Let us assume that we are interested in only two states, $|k\rangle $ and $|l\rangle $. In this case, we can approximate the Hamiltonian as $\mathcal{H}_{\mathrm{QA}}\simeq \frac{\Delta}{2}\sigma_{z}=|l\rangle \langle l| - |k\rangle \langle k|$.
Furthermore, we can use Eq.~\eqref{eq:qubit_rwahamil_mod} for the effective Hamiltonian.
The transition amplitude~\eqref{eq:traamp} is calculated as
\begin{widetext}
\begin{align}
    \braket{\phi(t)|e^{-irt\frac{\Delta}{2}\sigma_{z}} e^{-it\mathcal{H}_{\mathrm{eff}}}|\psi(0)}&=\cos{\frac{\tilde{\lambda}'}{2}t}\braket{\phi(t)|e^{-irt\frac{\Delta}{2}\sigma_{z}}|\psi(0)}-i\sin{\frac{\tilde{\lambda}'}{2}t}\braket{\phi(t)|e^{-irt\frac{\Delta}{2}\sigma_{z}}U^{\dag}_{\rm diag} \sigma_{x} U_{\rm diag}|\psi(0)},\nonumber\\
&=\cos{\frac{\tilde{\lambda}'}{2}t}\left(\braket{\phi(t)|l}\braket{l|\psi(0)}e^{-irt\frac{\Delta}{2}}+\braket{\phi(t)|k}\braket{k|\psi(0)}e^{irt\frac{\Delta}{2}}\right)\nonumber\\
&\quad -i\sin{\frac{\tilde{\lambda}'}{2}t}\left(\braket{\phi(t)|l}\braket{l|U^{\dag}_{\rm diag} \sigma_{x} U_{\rm diag}|\psi(0)}e^{-irt\frac{\Delta}{2}}+\braket{\phi(t)|k}\braket{k|U^{\dag}_{\rm diag} \sigma_{x} U_{\rm diag}|\psi(0)}e^{irt\frac{\Delta}{2}}\right).\label{eq:Rabiamp}
\end{align}
\end{widetext}

We can see that the absolute square of Eq.~\eqref{eq:Rabiamp} includes five different frequency modes:
\begin{align}
  \Omega &=  0,\ \tilde{\lambda}',\ r\Delta-\tilde{\lambda}',\ r\Delta,\ r\Delta+\tilde{\lambda}'.\nonumber\\
  &= 0,\ \tilde{\lambda}',\ \omega-\tilde{\lambda}',\ \omega,\ \omega+\tilde{\lambda}'.
\end{align}
Although we have five peaks, it is easy to specify the target peak for the following reason. As mentioned in the main text, we sweep the frequency range of
$0<\Omega \ll \omega $; hence, we observe only a peak $\tilde{\lambda}'$.

\section{Results for strong non-adiabatic transitions}
\label{sec:nonadiabatic}
In the main text, we considered a case in which the non-adiabatic transition is not relevant. In this section, we investigate the performance of our scheme when we increase the effect of the non-adiabatic transitions in case $E$.

\begin{figure}[ht]
    \begin{tabular}{c}
    \includegraphics[width=8cm]{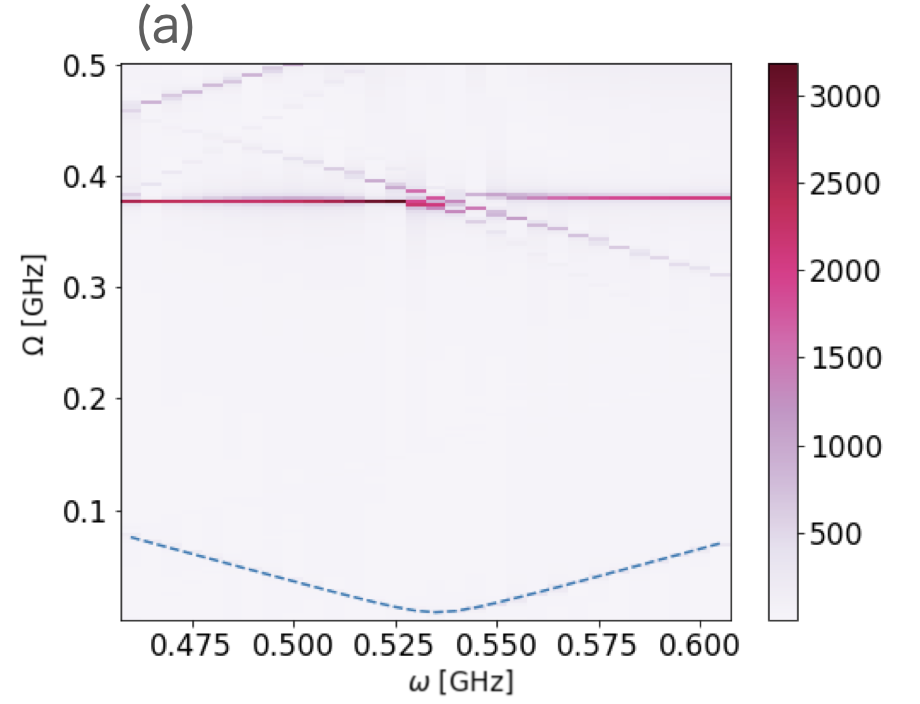}\\
    \includegraphics[width=8cm]{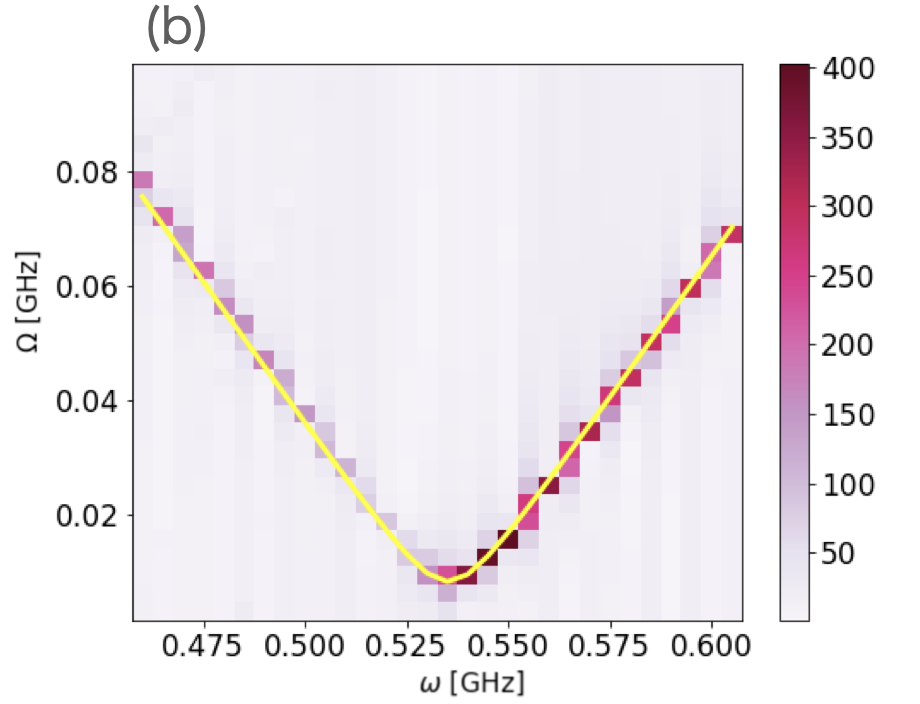}
    \end{tabular}
    \caption{In case E,
    we plot the power spectrum $P(\omega,\Omega)$, where we set $T_\mathrm{ann}=3$ and $t_{1}/T_\mathrm{ann}=0.9$ in (a). The magnified view is shown in (b).
    The blue and yellow lines represent the target peaks obtained by diagonalization. The peak we want is seen at the correct position when $\Omega$ is enlarged in the small region; however, it is lower than the high-frequency peak around $\Omega = 0.38$.
    }
    \label{fig:power-spectrum-npos1num4}
\end{figure}

We plot the spectrum by setting $T_\mathrm{ann} =3$ and $s_{1}= 0.9$,
as shown in Fig.~\ref{fig:power-spectrum-npos1num4} (a). Here, as a visual guide,
we plot a blue line corresponding to the analytical curve of Eq.~\eqref{eq:hyperbolic-curve}, where we use the actual values of $|\braket{1|\dot{\mathcal{H}}|0}|$ and $(E_{1}-E_{0})$, and this is the target peak in the spectrum. The magnified view is shown in Fig.~\ref{fig:power-spectrum-npos1num4} (b).

We observe unexpected peaks at frequencies of around $\Omega = 0.38$ in the spectrum, and their height is more significant than that of the target peak. Therefore, if we naively adopt our method described in Eq.~\eqref{def:omega_est}, we obtain incorrect estimated values of the transition matrix element and energy gap.

\begin{figure}
    \centering
    \includegraphics[width = 8cm]{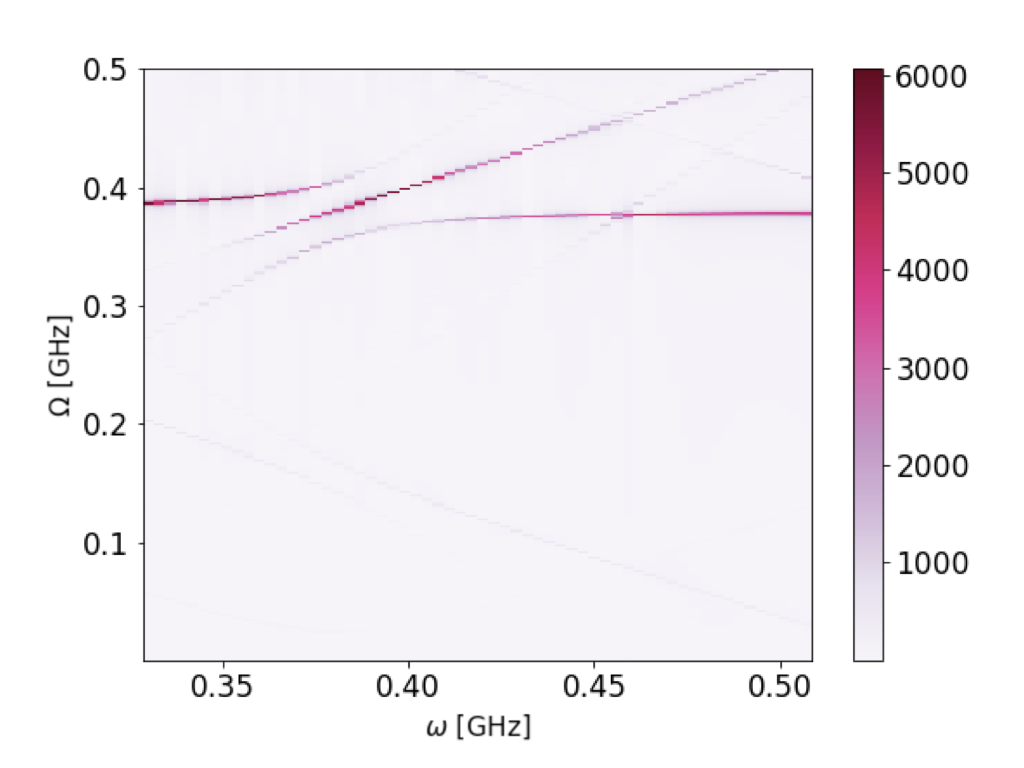}
    \caption{Plot of the spectrum to focus on the peak owing to a transition from the first excited state to the second excited stat, in case E. We observe three peaks around $\Omega \simeq \omega \simeq 0.38$.
    These correspond to frequencies of
    $\Omega = \omega - \Omega_{\mathrm{ana}}^{(1,2)}(\omega),\ \omega,\ \omega + \Omega_{\mathrm{ana}}^{(1,2)}(\omega)$.
    }
    \label{fig:triplet}
\end{figure}

To understand the origin of these unexpected peaks, we perform analytical calculations in order to obtain resonant frequencies in the spectrum with non-adiabatic transitions in Appendix~\ref{subsec:apb2-supposcase}.
Although the peaks at
$\Omega = \omega - \Omega_{\mathrm{ana}}^{(0,1)}(\omega),\ \omega,\ \omega + \Omega_{\mathrm{ana}}^{(0,1)}(\omega)$ should exist, we could not observe them owing to the restricted range of $\Omega$ in the spectrum, as mentioned in Appendix~\ref{subsec:apb2-supposcase}. Meanwhile,
if there is a non-negligible population of the second excited state induced by the non-adiabatic transitions,
we can observe peaks at frequencies of $\Omega = \omega - \Omega_{\mathrm{ana}}^{(1,2)}(\omega),\ \omega,\ \omega + \Omega_{\mathrm{ana}}^{(1,2)}(\omega)$.
To see these points, we plot the spectrum in Fig.~\ref{fig:triplet}, and we actually observe three such peaks.
Moreover, when $\omega$ is far from the energy difference $(E_{2}-E_{1})$, the peak at a frequency of $\omega - \Omega_{\mathrm{ana}}^{(1,2)}(\omega)$ asymptotically approaches the energy difference $(E_{2}- E_{1})$, and this is the origin of the highest peaks in Fig. \ref{fig:power-spectrum-npos1num4}(a).
The frequency of $\omega - \Omega_{\mathrm{ana}}^{(1,2)}(\omega)$ is much lower than
$\omega - \Omega_{\mathrm{ana}}^{(0,1)}(\omega),\ \omega,\ \omega + \Omega_{\mathrm{ana}}^{(0,1)}(\omega)$; therefore, we  observe this peak even if we restrict the range of $\Omega$ in the spectrum.

Even when we observe peaks other than the target peaks, there is a way to estimate the transition matrix element and energy gap.
As mentioned in the main text (in Section \ref{subsubsec:case_E}), we adopt a modified method to identify the target peak, which is useful for this case as well.
The positions of the target peak are expected to be fitted by the analytical formula in Eq.~\eqref{eq:hyperbolic-curve}. Thus, if we fail to fit the peaks, we can guess that such peaks do not correspond to the peak from the Rabi oscillation. Indeed, the peak $\Omega\simeq 0.38$ cannot be well fitted by Eq.~\eqref{eq:hyperbolic-curve}.
Meanwhile, if we focus on the peaks with frequencies of around $\Omega=0.01$, as shown in Fig.~\ref{fig:power-spectrum-npos1num4} (b), we can fit these peaks by the analytical formula; hence, we can accurately estimate the transition matrix element and energy gap.

\begin{figure}[ht]
    \begin{tabular}{c}
    \includegraphics[width=8cm]{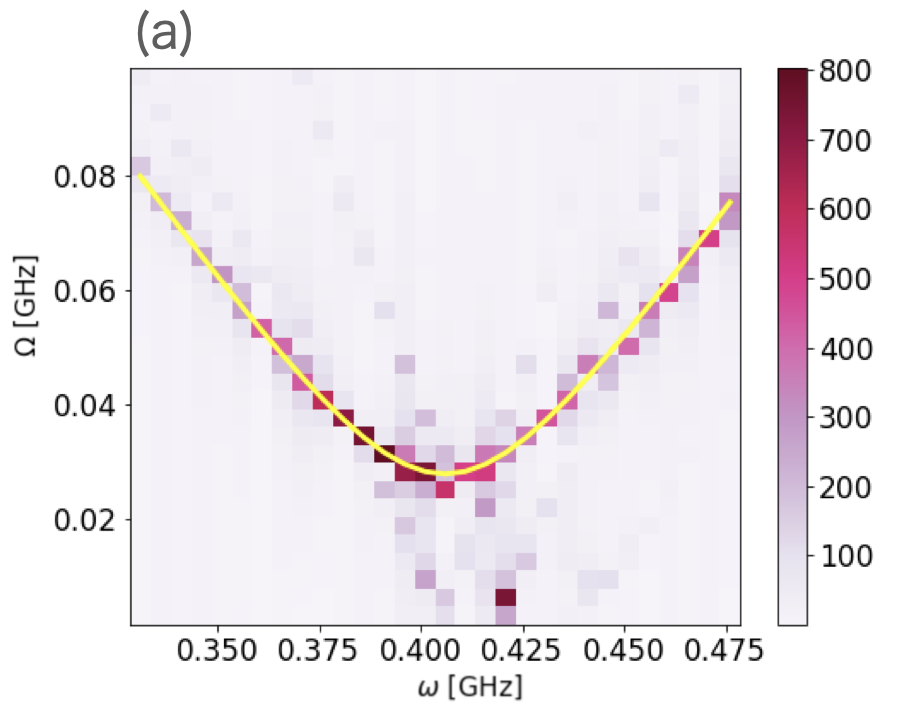}\\
    \includegraphics[width=8cm]{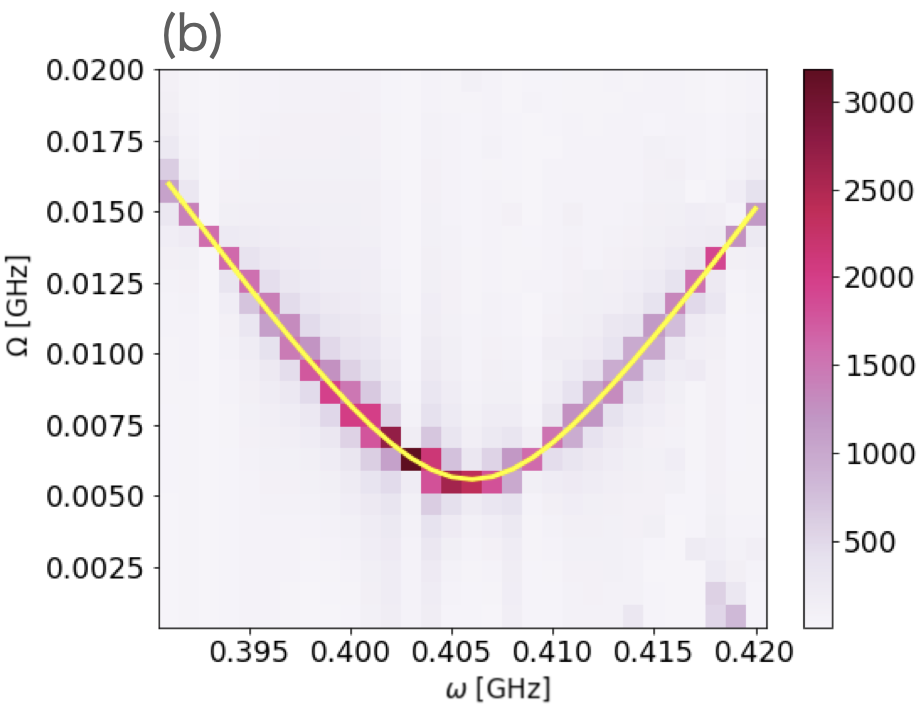}
    \end{tabular}
     \caption{Plot of the power spectrum in case E with $T_{\rm ann}=3$ and $t_{1}/T_{\rm ann}=0.7$. Here, as a visual guide, we plot a yellow line corresponding to the analytical curve of Eq.~\eqref{eq:hyperbolic-curve}, which are the target peaks. (a) We select $\lambda/T_{\rm ann} = 0.05$. We observe the target peaks as well as the peaks due to higher-order perturbations.
     (b) We select $\lambda/T_\mathrm{ann}=0.01$. Compared to the case with $\lambda/T_\mathrm{ann}=0.05$, the peaks caused by higher-order perturbations are smaller.}
    \label{fig:power-spectrum-npos1num3-add}
\end{figure}

Furthermore, we plot the spectrum by setting $T_\mathrm{ann} =3$ and $s_{1}= 0.7$
in Fig.~\ref{fig:power-spectrum-npos1num3-add} (a). Here, as a visual guide, we plot a yellow line corresponding to the analytical curve of Eq.~\eqref{eq:hyperbolic-curve}, which are the target peaks. Here, the target peaks as well as other peaks are observed.
These come from higher-order perturbations, which can be observed for a larger Rabi frequency (see Appendix \ref{sec:higher-pert}).
We can distinguish these secondary peaks from the target peaks as follows.

First, the secondary peaks are usually smaller than the target peaks. As shown in Fig.~\ref{fig:power-spectrum-npos1num3-add}, except for a few points, the target peaks are the highest in this frequency region. Second, from the fitting results by Eq.~\eqref{eq:hyperbolic-curve}, we can distinguish the target peaks from the secondary peaks (for example, the slope of the target peaks is given by
$\frac{d\Omega_{\mathrm{ana}}(\omega)}{d\omega}\simeq 1$ for large $\omega$, while the slope of the secondary peaks is twice as large).
Third, the $\lambda$-dependence of the peak height is different.
The height of the target peaks scales as $\lambda ^{2}$, while that of the secondary peaks scales as $\lambda^{4}$, and this lets us identify the target peaks by sweeping $\lambda$.
As shown in Fig.~\ref{fig:power-spectrum-npos1num3-add}, we plot the spectrum by selecting a smaller $\lambda$, and we show that the secondary peak becomes nearly invisible compared to the target peak.

\section{Perturbative approach of this method}
\label{sec:higher-pert}

In the main text, we used RWA, and the Hamiltonian was effectively transformed into a simple two-dimensional one. When we consider higher-order perturbations, there can be other peaks without the $|E_{1}-E_{0}|=\omega$ condition.

We calculate the transition amplitude $\braket{f|U(t,0)|i}$, where $U(t,0)$ is a unitary operator expressing the time evolution from time $0$ to time $t$, and $\ket{i}$ and $\ket{f}$ are the initial state and the final state, respectively.

First, we describe the Hamiltonian of Eq.~\eqref{def:external_Hamiltonian} in the so-called interaction picture:
\begin{align}
    \tilde{\mathcal{H}}(t) = \lambda e^{it\mathcal{H}_{\mathrm{QA}}}\dot{\mathcal{H}}_{\rm QA}e^{-it\mathcal{H}_{\mathrm{QA}}}\cos{\omega t}.\label{eq:interaction-hamiltonian-pert}
\end{align}
Then, the transition amplitude is given by
\begin{align}
    &\braket{f|U(t,0)|i} = \braket{f|i}+(-i)\int_{0}^{t}d\tau \braket{\tilde{f}|\tilde{\mathcal{H}}(\tau)|\tilde{i}}\nonumber\\
    &\qquad+(-i)^{2}\int_{0}^{t}d\tau_{1}\int_{0}^{\tau_{1}}d\tau_{2}\braket{\tilde{f}|\tilde{\mathcal{H}}(\tau_{1})\tilde{\mathcal{H}}(\tau_{2})|\tilde{i}}+\cdots, \label{eq:Dyson-series}
\end{align}
where the tilde symbol implies that the states are in the interaction picture.
As the Schroedinger picture coincides with the interaction picture at $t=0$, we have $|\tilde{i}\rangle =|i\rangle $.
We assume that $\lambda$ is small, and we calculate the transition amplitude up to the second order of the perturbation theory.

In our method,
we assume that the dynamics is adiabatic in the second and fourth steps.
In this case, we can set $|i\rangle =|0\rangle $ and $|f\rangle =|m\rangle $, and we evaluate the quantity of $p_{0,m}(t) = |\braket{m|U(t,0)|0}|^{2}$.

\subsection{First-order perturbation and the Rabi oscillation}

Let us consider the first order of the perturbative term in Eq. \eqref{eq:Dyson-series} as follows:
\begin{align}
    &\int_{0}^{t} d\tau\braket{\tilde{m}|\tilde{\mathcal{H}}(\tau)|\tilde{0}}\nonumber\\
    &=\lambda\int_{0}^{t} d\tau\braket{m|e^{-it\mathcal{H}_{\mathrm{QA}}}e^{i\tau\mathcal{H}_{\mathrm{QA}}}\dot{\mathcal{H}}_{\rm QA}e^{-i\tau\mathcal{H}_{\mathrm{QA}}}|0}\cos\omega\tau\nonumber\\
    &=\lambda\braket{m|\dot{\mathcal{H}}_{\rm QA}|0}\int_{0}^{t}d\tau e^{-iE_{m}t+i(E_{m}-E_{0})\tau}\cos{\omega\tau}\nonumber\\
    &=\frac{\lambda}{2}\braket{m|\dot{\mathcal{H}}_{\rm QA}|0}e^{-iE_{m}t}\nonumber\\
    &\qquad\cdot\left(\frac{e^{i(E_{m}-E_{0}+\omega)t}-1}{i(E_{m}-E_{0}+\omega)}+\frac{e^{i(E_{m}-E_{0}-\omega)t}-1}{i(E_{m}-E_{0}-\omega)}\right).\label{eq:perturb-first}
\end{align}
The absolute square of Eq.~\eqref{eq:perturb-first} includes
terms with frequencies of $2\omega$, $E_{m}-E_{0}\pm\omega$.
This result is consistent with the analytical result in Eq.~\eqref{eq:hyperbolic-curve} in the limit of small $\lambda$, which is used to predict the resonance at $\Omega=|E_{m}-E_{0}-\omega|$.

\subsection{Second-order perturbation}

Let us consider three energy eigenstates of $\mathcal{H}_{\mathrm{QA}}$ as $\ket{0}, \ket{I}, \ket{m}$, and we assume that $\ket{i} =\ket{0}$ and $\ket{f} = \ket{m}$. The third term of the right-hand side of Eq. \eqref{eq:Dyson-series} is given by
\begin{align}
    &(-i)^{2}\lambda^{2}\int_{0}^{t} dt_1\int_{0}^{t_1}dt_{2}\sum_{i} \braket{m|\dot{\mathcal{H}}_{\rm QA}|i}\braket{i|\dot{\mathcal{H}}_{\rm QA}|0}\nonumber\\
    &\quad\quad \cdot e^{-iE_{m}t+i(E_{m}-E_{i})t_{1}} e^{i(E_{i}-E_{0})t_{2}}\cos{\omega t_{1}}\cos{\omega t_{2}},\\
    &=(-i)^{2}\lambda^{2}\braket{m|\dot{\mathcal{H}}_{\rm QA}|I}\braket{I|\dot{\mathcal{H}}_{\rm QA}|0}\nonumber \\
    &\qquad \times\frac{e^{i(E_{m}-E_{0}-2\omega)t}-1}{(E_{I}-E_{0}-\omega)(E_{m}-E_{0}-2\omega)}\nonumber \\
    &\quad+ \cdots.\label{square-pert}
 \end{align}
The absolute square of this amplitude includes various modes; however, one of them is $E_{m}-E_{0}-2\omega$. Thus, the probability function $p_{0,m}(\tau)$ includes an oscillation with a frequency of $\Omega=2(\omega-(E_{m}-E_{0})/2)$.
Importantly, for this frequency, we have $\frac{d \Omega}{d\omega}=2$, while we have $\frac{d \Omega_{\mathrm{ana}}}{d\omega}=1$ for our analytical formula in Eq.~\eqref{eq:hyperbolic-curve}.

\section{On the validity of the adiabatic condition~\eqref{eq:adiabatic_criterion}}
\label{sec:valid}
\begin{figure}
\includegraphics[width = 8cm]{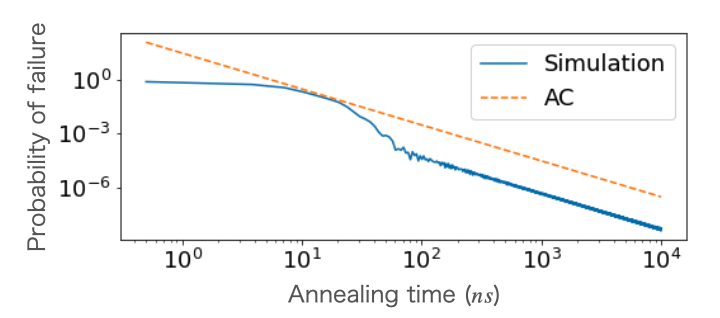}
\caption{Plot of the population of being one of the excited states (solid line) and sum of $p_{n}(1)$ (Dotted line) against the annealing time $T_{\rm ann}$ for $n= 1,2,3$. 
For the large $T_{\rm ann}$, we confirm that $p_{n}(1)$ actually
gives an upper bound of the population in these cases.}\label{figure:popl}
\end{figure}

The adiabatic condition~\eqref{eq:adiabatic_criterion} is believed to be related to success of quantum annealing but it is not proven to be sufficient. 
The adiabatic condition Eq.~\eqref{eq:adiabatic_criterion} comes from the equation~\eqref{eq:anm}, and the final transition amplitude to the excited states is approximately given by~\eqref{eq:approx_wavf}.
It is worth mentioning that, since a perturbative theory is used, the Eq.~\eqref{eq:approx_wavf} is not exact.
Although we cannot show the validity of the Eq.~\eqref{eq:approx_wavf} for any quantum annealing, we can show that the Eq.~\eqref{eq:approx_wavf} actually provides us with an upperbound of the population of the excited states for our numerical examples  in section~\ref{sec:Numcal}.
According to Eq.~\eqref{eq:approx_wavf}, the population of the $n$-th excited state is evaluated by
\begin{align}
    |c_{n}(s)|^{2} &= |A_{0n}(s)-A_{0n}(0)|^{2}\nonumber \\
    &\leq \left(\max_{0\leq\sigma\leq s}[|A_{0n}(\sigma)|]+|A_{0n}(0)|\right)^{2} =: p_{n}(s).\label{def:pn}
\end{align}
We perform a numerical simulation of the quantum annealing defined by~\eqref{def:QA_Hamiltonian_conv} with the two qubit model~\eqref{eq:twoqubit-Hamiltonian}, and calculate the populations of the first, the second, and the third excited states at the time $s=1$. Then, we plot the dependence of these populations on the annealing time $T_{\rm ann}$. (see Fig.~\ref{figure:popl})

Fig.~\ref{figure:popl} shows that sum of $p_{n}$ defined by~\eqref{def:pn} is actually larger than the populations obtained by the quantum annealing for the large $T_{\rm ann}$. Also, we confirm that the populations are inversely proportional to the square of the annealing time $T_{\rm ann}$ in the limit of the large $T_{\rm ann}$. 
If $T_{\rm ann}$ is small, the approximation used in~\eqref{eq:integral_equation3} is violated and $p_{n}$ does not provide the upper bound of the population. 
Thus, the adiabatic condition we used~\eqref{eq:adiabatic_criterion} is expected to be valid for the large $T_{\rm ann}$.

\section{Numerical results for larger systems}
\label{more_complex}

To establish the applicability of our method for larger systems, we adopt a nine-qubit system with the following Hamiltonian.
\begin{align}
\mathcal{H}_\mathrm{D}&=\sum_{i=1}^{9}h_{i}\sigma_{z}^{i},\nonumber\\ \mathcal{H}_\mathrm{P}&=\left(\sum_{i=1}^{8}J_{i,i+1}\sigma_{x}^{i}\sigma_{x}^{i+1}\right)+J_{9,1}\sigma^{9}_{x}\sigma^{1}_{x}\label{Ham_9bit}, 
\end{align}
where $\sigma_{x}^{i}$ ($\sigma_{z}^{i}$) is the Pauli $X$ ($Z$) matrix acting on $i$-th qubit, $J_{i,i+1}$ is a coupling strength between nearest neighbor qubits, $h_{i}$ is a strength of the external field. Actual values of these parameters are shown in Table~\ref{tab:params}.

\begin{table}[]
    \centering
    \caption{Parameters of the nine-qubit simulation.}\label{tab:params}
    \begin{tabular}{ccccccccc}
    \hline
       $h_{1}$ & $h_{2}$ & $h_{3}$ & $h_{4}$ & $h_{5}$ & $h_{6}$ & $h_{7}$ & $h_{8}$ &$h_{9}$\\
       1.00 & 0.10  &0.75 &0.60 &1.10 &1.30 &0.30 &0.26 &1.42\\ \hline
       $J_{1,2}$&  $J_{2,3}$ & $J_{3,4}$ & $J_{4,5}$ & $J_{5,6}$ & $J_{6,7}$ & $J_{7,8}$ &$J_{8,9}$ & $J_{9,1}$\\
       -0.91&-0.58 & -1.04 & -0.95 &-0.93&-0.98 &-0.75 &-0.88 &-1.01\\
       \hline
    \end{tabular}
\end{table}

\begin{figure}
    \includegraphics[width = 8.5cm]{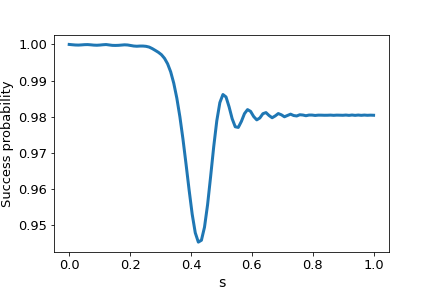}
    \caption{Success probability of our nine-qubit annealing process with respect to $s$. Around $s=0.4$, success probability is suddenly worsen.}\label{fid_9qb}
\end{figure}

\begin{figure}
    \centering
    \includegraphics[width = 9cm]{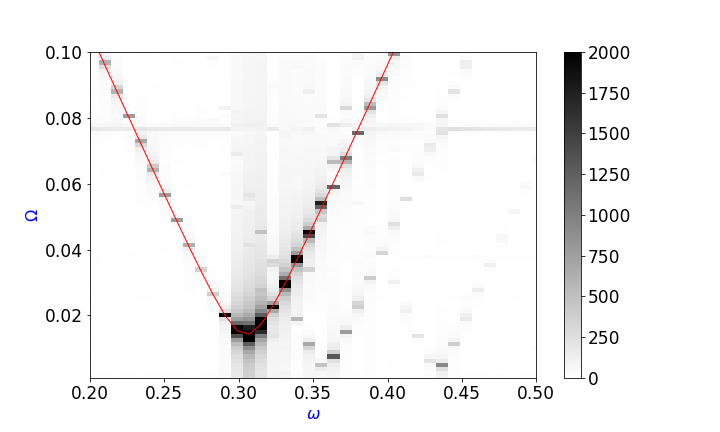}
    \caption{Power spectrum for measuring of $E_{4}-E_{0}$ and $\langle 4|\dot{H}|0\rangle$ with our method. The brown curve shows the analytically expected curve $\Omega_{\rm ana}(\omega)$ 
    in Eq.~\eqref{eq:hyperbolic-curve} 
    with diagonalization of the Hamiltonian Eq.~\eqref{Ham_9bit}. We select $\lambda/T_{\rm ann} = 0.01$ in this simulation. }\label{spectrum_9qb}
\end{figure}

Let us explain how we can apply our method to this system.
During the implementation of QA with the driver Hamiltonian and problem Hamiltonian described above, 
a notable observation emerges where
the success probability suddenly decreases at a specific point in the process. (see Fig.~\ref{fid_9qb}). 
We apply our method to measure the adiabatic condition at the point, and we obtain a power spectrum depicted as Fig.~\ref{spectrum_9qb}.
As depicted in Fig.~\ref{spectrum_9qb}, while the spectrum displays numerous curves, the curve corresponding to the desired signal is discernible, and so we can fit this by using the analytical expression of Eq.~\eqref{eq:hyperbolic-curve}.
Therefore, we can estimate the value of the right-hand side of the Eq.~\eqref{eq:adiabatic_criterion} from the spectrum, and this information is valuable when we try to optimize the schedule of QA to maximize the success probability.

\bibliography{main}

\end{document}